\documentclass[superscriptaddress,showpacs,amssymb,10pt,reprint,aps,prd,longbibliography,nofootinbib,floatfix]{revtex4-1}
\usepackage{graphicx,epsfig,amssymb} 
\usepackage{amsmath,amsfonts, times}
\usepackage{bm} 
\usepackage[normalem]{ulem}
\usepackage{epstopdf}
\usepackage[linktocpage,colorlinks]{hyperref}
\usepackage[caption=false]{subfig}
\usepackage[usenames]{color} 
\usepackage{mathrsfs} 
\usepackage{natbib}
\usepackage{soul}
\usepackage{subfig}
\usepackage[utf8x]{inputenc}
\usepackage{tikz}
\usetikzlibrary{decorations.pathmorphing}
\definecolor{coolblack}{rgb}{0.0, 0.18, 0.39}
\definecolor{darkred}{rgb}{0.5,0,0}
\definecolor{darkgreen}{rgb}{0,0.5,0}
\definecolor{darkblue}{rgb}{0,0,0.5}
\definecolor{lapislazuli}{rgb}{0.15, 0.38, 0.61}
\definecolor{venetianred}{rgb}{0.78, 0.03, 0.08}
\definecolor{bleudefrance}{rgb}{0.19, 0.55, 0.91}
\definecolor{dogwoodrose}{rgb}{0.84, 0.09, 0.41}
\hypersetup{colorlinks=true, citecolor=darkgreen, linkcolor=darkblue, 
urlcolor = blue}

\newcommand\numberthis{\addtocounter{equation}{1}\tag{\theequation}}

\begin{document}
	\title{\large Wormholes in Lorentz-violating gravity}
	\author{Renan B. Magalh\~aes}
    \email{renan.batalha@ufma.br}
	\affiliation{Programa de Pós-graduação em Física, Universidade Federal do Maranhão, Campus Universitário do Bacanga, São Luís (MA), 65080-805, Brazil.}
\author{Leandro A. Lessa}
	\email{leandrophys@gmail.com}
	\affiliation{Programa de Pós-graduação em Física, Universidade Federal do Maranhão, Campus Universitário do Bacanga, São Luís (MA), 65080-805, Brazil.}
\author{Manoel M. Ferreira Jr.}
	\email{manoel.messias@ufma.br}
	\affiliation{Programa de Pós-graduação em Física, Universidade Federal do Maranhão, Campus Universitário do Bacanga, São Luís (MA), 65080-805, Brazil.}
	\affiliation{Departamento de Física, Universidade Federal do Maranhão, Campus Universitário do Bacanga, São Luís, Maranhão 65080-805, Brazil}
\begin{abstract}
We investigate the possibility of obtaining traversable wormholes supported by phantom scalar fields in Lorentz-violating gravity with an antisymmetric rank-2 tensor with a non-zero vacuum expectation value non-minimally coupled to the curvature tensor. This Lorentz violation framework shows to be a suitable scenario to search for wormhole solutions in the presence of Lorentz violation, since it introduces mild constraints on the areal radius. The vacuum expectation value of the antisymmetric rank-2 tensor, nonetheless, imposes constraints on the lapse function. As a consequence, under the vacuum configuration adopted, the allowed lapse functions can be either constant, linear or quadratic, depending on the self-interaction potential that drives the spontaneous breaking of the Lorentz symmetry. Thus, we find the Ellis-Bronnikov counterpart in this Lorentz-violating scenario as well as Lorentz-violating wormholes with a Rindler-type acceleration and an effective cosmological constant. Properties of these wormholes, such as their non-flat asymptotics, are investigated. 

\end{abstract}
\date{\today}
\maketitle
\section{Introduction}
Unifying the description of gravity and quantum mechanics within a single theoretical framework remains a profound challenge. Although a reasonably satisfactory theory of linearized perturbative quantum gravity can be formulated on a fixed background spacetime, significant difficulties arise in developing a non-perturbative theory of quantum gravity~\cite{Page:1981aj,eppley1977necessity,penrose2014gravitization}, such as Canonical Quantum Gravity~\cite{isham1993canonical}. Coupling General Relativity (GR) with the Standard Model (SM) of particle physics results in an incomplete quantum framework. Nonetheless, this combination aligns remarkably well with experimental data in appropriate regimes~\cite{Hawking:1973uf,Donoghue:1994dn}. Consequently, any deviations from established physics arising from an underlying unified theory are expected to be small, likely suppressed by a large energy scale, such as the Planck mass. Nevertheless, progress can still be made through effective field theory (EFT), which offers a powerful and model-independent framework for studying these small deviations~\cite{weinberg2016effective}.

An intriguing possibility regarding observable effects of an underlying quantum gravitational theory is the emergence of small residual couplings at the level of EFT~\cite{ghosh2023does,Addazi:2021xuf,liang2022polarizations,Gupta:2024gun}. These coupling coefficients can vary with spacetime position, assuming the underlying theory generates solutions with non-trivial backgrounds. Consequently, such background coefficients are better represented as tensors~\cite{kostelecky2021backgrounds}. Notably, non-constant background tensors serving as couplings in EFT imply violations of local Lorentz invariance~\cite{Kostelecky:2020hbb}. The framework for the general EFT based on GR coupled to the SM, incorporating coefficients that lead to violations of local Lorentz invariance, is detailed in Ref.~\cite{Kostelecky:2003fs}. 

In particular, a mechanism for local Lorentz violation (LV) is provided by a spontaneous symmetry breaking potential due to self-interacting tensor fields. The vacuum expectation value (VEV) of these tensor fields generates background tensors which, when coupled to SM, violate the local Lorentz symmetry \cite{Kostelecky:2003fs,Lessa:2023yvw}. Moreover, the spontaneous violation of local Lorentz invariance also implies spontaneous violation of diffeomorphism invariance~\cite{Bluhm:2004ep}. It is worth emphasizing that, in Lorentz-violating theories, the presence of non-zero VEV of tensor fields implies the existence of at least two possible classes of transformations: observer transformations and particle transformations~\cite{colladay1997cpt}. The former involves changes in the observer's frame. Observer coordinate transformations on the manifold are general coordinate transformations (observer diffeomorphism), which leave the action invariant. Therefore, it is expected that any realistic physical theory should be covariant under these transformations. On the other hand, particle transformations are defined to act on individual particles, while leaving unchanged VEV. Similarly, particle diffeomorphisms with incorporated pullback leave all vacuum values unaffected. Thus, the violation of local Lorentz symmetry and diffeomorphism symmetry are related to particle transformations. Furthermore, spontaneous LV allows the Lorentz-violating terms in the Lagrangian to satisfy the Bianchi identities, a key property for the gravitational field~\cite{Kostelecky:2003fs}.

For instance, the simplest field exhibiting these characteristics is the self-interacting vector field, known as the bumblebee field. This field has been extensively studied in the literature, both in flat~\cite{Hernaski:2014jsa,Maluf:2015hda,Seifert:2009gi} and curved spacetimes~\cite{Casana:2017jkc,Filho:2022yrk,Maluf:2020kgf,Maluf:2021lwh,Lessa:2023dbd,araujo2024exact}. Furthermore, the spontaneous breaking of Lorentz symmetry leads to the emergence of Nambu-Goldstone (NG) modes and massive modes~\cite{Bluhm:2007bd}. For a quadratic potential, within the so-called Kostelecky-Samuel model in \(3+1\) dimensions, fluctuations around the VEV produce two transverse NG modes and one longitudinal massive mode \cite{Bluhm:2004ep,Bluhm:2007bd}. Another well-studied Lorentz-violating field theory involves an antisymmetric rank-2 tensor coupled to gravity, also known as Kalb-Ramond fields \cite{PhysRevD.9.2273}, denoted as $B_{\mu\nu}=-B_{\nu\mu}$. Unlike the bumblebee field, the VEV of the background field $B_{\mu\nu}$ can be decomposed into two spatial vectors and one temporal vector, analogous to the decomposition of the electromagnetic tensor~\cite{Lessa:2019bgi}. This allows greater flexibility in selecting the preferred orientation these fields can exhibit. In the context of gravitational solutions generated by the VEV of this field, one finds black hole solutions \cite{Yang:2023wtu,Duan:2023gng,Liu:2024oas}, wormholes~\cite{Maluf:2021ywn}, and cosmological solutions~\cite{Maluf:2021eyu}. A broader analysis of this field in Lorentz-violating frameworks is detailed in Ref.~\cite{Altschul:2009ae}.

Lorentz-violating extension of GR are constructed using Riemann-Cartan geometry. In the Riemannian limit of the theory, the torsion tensor vanishes. The leading-order Lagrangian terms for this zero-torsion theory consist in the usual Einstein-Hilbert term and possible cosmological constant, along with the dominant curvature coupling terms involving LV. These are governed by three coefficient fields, denoted by \( u \), \( s^{\mu\nu} \), and \( t^{\mu\nu\alpha\beta} \). The coupling of these coefficients with curvature tensors gives rise to non-minimal models.  In such models, the couplings between the VEV of background fields and the curvature introduce constraints on the spacetime geometry -- hereafter called VEV equations -- that stem from the spacetime anisotropy induced by LV. These constraints should be verified alongside the gravitational field equations and the equations governing matter fields. Notably, the antisymmetric 2-tensor field has the advantage of generating possible non-minimal curvature coupling terms with \( t^{\mu\nu\alpha\beta} \), unlike the case of the bumblebee field, which only allows couplings with the coefficients \( u \) and \( s^{\mu\nu} \) \cite{Kostelecky:2003fs}. Moreover, the VEV coefficient  $ \langle t^{\mu\nu\alpha\beta} \rangle$  is absent at leading orders in the post-Newtonian expansion, a phenomenon referred to in the literature as the `t puzzle'~\cite{Bonder:2015maa}. However, nothing prevents the emergence of gravitational corrections beyond this limit. Indeed, as we shall see throughout this paper, it is possible to obtain wormhole solutions involving this type of Lorentz-violating curvature coefficient. 

Although wormholes are often regarded as exotic geometric objects, they have attracted growing interest in theoretical physics due to their potential observational signatures~\cite{deyGravitationalLensingWormholes2008,ohgamiWormholeShadows2015,cardosoGravitationalWaveRingdownProbe2016,guerreroLightRingImages2022,furutaSpectralLinesDirty2024,magalhaesAsymmetricWormholesPalatini2023,magalhaesEchoesBoundedUniverses2024}. Significant efforts have been devoted to understanding and modeling their properties, as well as exploring their detectable features~\cite{simonettiSensitiveSearchesWormholes2021,daiObservingWormhole2019}. These objects with non-trivial topology may connect either distinct universes or serve as a conduit within the same universe~\cite{Visser:1995cc}. Within the framework of GR, to ensure their traversability typically requires the violation of energy conditions~\cite{morrisWormholesSpacetimeTheir1988,sharmaGeneralisedEllisBronnikov2021}. For instance, Ellis-Bronnikov (EB) solution~\cite{ellisEtherFlowDrainhole1973,bronnikovScalartensorTheoryScalar1973}, one of the simplest wormhole solutions, are supported by phantom scalar fields, characterized by a "wrong'' sign in front of its kinetic term. This feature inherently leads to the violation of some of these energy conditions. Wormholes arise as geometrical entities also in modified theories of gravity~\cite{Bambi:2015zch,Capozziello:2012hr,Rosa:2018jwp,Bronnikov:2019ugl,magalhaesCompactObjectsQuadratic2022}. In the context of modified gravitational models that violate Lorentz symmetry, spontaneous breaking of local Lorentz symmetry may play a notable role in the generation of wormholes~\cite{Maluf:2021ywn}.

The aim of this work is to explore phantom wormhole solutions generated by non-minimal models involving spontaneous Lorentz symmetry breaking triggered by a rank-2 antisymmetric background VEV field. The model incorporates non-minimal couplings between the VEV and the Riemann tensor, which proved to be a suitable framework to generate wormhole configurations. Assuming a spacetime with static and spherical symmetry, we identified four distinct classes of Lorentz-violating wormhole solutions with different asymptotic regimes. A fundamental part of our analysis lies in the VEV equation of the antisymmetric rank-2 tensor that, under the vacuum configuration chosen in this work, imposes geometric constraints solely on the lapse function and keeps the areal radius unconstrained. It is important to point out that the satisfaction of this constraint is required to obtain self-consistent wormhole solutions within this Lorentz-violating framework~\cite{Lessa:2025kln}.
This mild constraint is used to determine the allowed lapse function, which can be, depending on the form of the self-interaction potential of the antisymmetric rank-2 tensor, either constant, linear or quadratic. Moreover, it is noteworthy that the asymptotic region of the simplest wormhole found in this work, a counterpart of the EB wormhole in LV models triggered by non-zero VEV of an antisymmetric rank-2 tensor, exhibits a non-flat asymptotic, with its equatorial plane having a cone geometry with angle deficit (or excess).

The content of this paper is organized as follows. In Sec.~\ref{sec2}, we introduce the LV framework we will use, and derive its field equations. The wormhole solutions supported by phantom scalar field in the considered LV scenario are presented in Sec.~\ref{seciv}. In Sec.~\ref{secv}, we investigate some properties of the solutions obtained, such as their equatorial plane geometry through embedding diagrams, their curvature invariants, and the geodesic structure of massless particles. Finally, we summarize our results and discuss some perspectives in Sec.~\ref{con}.
\section{Framework}\label{sec2}
Let us consider the gravity theory described by the action
\begin{equation} \label{ac1}
    S = \int d^4x\sqrt{-g}\left\{ \frac{R}{2\kappa} +\frac{1}{2}\partial_{\mu}\phi\partial^{\mu}\phi - \mathcal{V}(\phi)   \right\}  +S_{LV},
\end{equation}
where $R$ is the Ricci scalar, $\phi$ is a phantom scalar field and $\mathcal{V}(\phi)$ its self-interaction potential. Furthermore, $S_{LV}$ encodes dynamical contributions of the Lorentz violation, through the Lorentz-violating coefficients, and it is given by \footnote{The general construction of the LV coefficients $u$, $s^{\mu\nu}$, and $t^{\mu\nu\alpha\beta}$ in terms of the field $B^{\mu\nu}$ can be found in Ref. \cite{Altschul:2009ae}.}
\begin{align*} \label{ac2}
  S_{LV} = &\int d^{4}x\sqrt{-g}  \bigg[ - \frac{1}{12}H_{\lambda\mu\nu}H^{\lambda\mu\nu} - V(X) \\&+ \frac{\xi}{2\kappa}B^{\mu\nu}B^{\alpha\beta}R_{\mu\nu\alpha\beta} \bigg],\numberthis
\end{align*}
where $\xi$ is an arbitrary constant. The field that carries the effects of Lorentz violation is a rank-2 antisymmetric tensor, denoted by the 2-form \( \boldsymbol{B}_2 = \frac{1}{2} B_{\mu \nu} \, dx^{\mu} \wedge dx^{\nu} \). Its field strength is a 3-form, given by \( \boldsymbol{H}_3 = d\boldsymbol{B}_2 \), where its components are
\begin{equation}
    H_{\lambda \mu \nu} = \partial_{\lambda} B_{\mu \nu} + \partial_{\mu} B_{\nu \lambda} + \partial_{\nu} B_{\lambda \mu}.
\end{equation}
Additionally, $ X \equiv B^{\mu\nu}B_{\mu\nu} + b^2$, with \( b^2 \in \mathbb{R}\).
The term of the self-interaction potential, $V$, triggers spontaneous breaking of the Lorentz symmetry. Indeed, \( V \) leads to the formation of a non-zero VEV for the 2-form field, i.e., 
\begin{equation} \label{vev1}
    \langle B_{\mu \nu} \rangle = b_{\mu \nu},
\end{equation}
which defines a background tensor field. Consequently, we have a local breaking of both Lorentz symmetry and diffeomorphism symmetry.

The VEV configuration (\ref{vev1}) is particularly interesting since it provides a simple framework for investigating the effects of LV in the gravitational sector. The potential \( V \), which drives the spontaneous breaking of Lorentz symmetry, takes a simple form in this configuration. For instance, we can consider the well-known smooth quadratic potential~\cite{Bluhm:2007bd} given by  
\begin{equation}   \label{quadr}
V^{Q} = \frac{1}{2} \zeta X^2,
\end{equation}  
where $\zeta$ is a constant.  
To ensure that (\ref{vev1}) represents a vacuum configuration, the constant norm condition, expressed as 
\begin{equation} \label{norm}
   - b^2 = \langle B_{\mu\nu}B^{\mu\nu} \rangle =  g^{\alpha\mu} g^{\beta\nu}   b_{\alpha\beta}b_{\mu\nu},
\end{equation}
 must be satisfied. Consequentently, the VEV $b_{\mu\nu}$ arises as a solution of $\langle V^Q \rangle=\langle V^{Q}_X \rangle=0$, where $V_X = \partial V/\partial X$. The smooth quadratic potential (\ref{quadr}) is particularly interesting once it generates both massless NG modes and massive modes~\cite{Bluhm:2004ep,Bluhm:2007bd}.  
 
 Alternatively, a linear potential in \( X \) can be constructed,  expressed as
 \begin{equation}   \label{line}
V^{L} = \lambda X,
\end{equation} 
where \( \lambda \) represents the Lagrange multiplier field~\cite{Bluhm:2004ep} . The equation of motion for \( \lambda \) enforces the vacuum condition \( X = 0 \), which consequently leads to \( V = 0 \) for any on-shell configuration of \( \lambda \).  Thus, we have $ \langle V^L \rangle =0$ and $\langle V^L_{X} \rangle =\langle\lambda\rangle$. As pointed out in Ref.~\cite{Maluf:2020kgf}, the VEV of the Lagrange multiplier can vary in spacetime. However, for the purposes of this work, we assume that $\langle\lambda\rangle\equiv \lambda$ is a real constant. Moreover, since the first-order derivative of the potential does not vanish in the VEV configuration, this potential is considered a candidate for the cosmological constant~\cite{Maluf:2020kgf}.

The gravitational field equations follow from the variation of the action (\ref{ac1}) with respect to the metric, namely
\begin{equation}
   \label{eq:einstein_KR}G_{\mu\nu} = R_{\mu\nu}-\frac{1}{2}R g_{\mu\nu} = \kappa T_{\mu\nu},
\end{equation}
where $T_{\mu\nu}= T_{\mu\nu}^{\phi}+T_{\mu\nu}^{LV}$ is the (total) energy momentum tensor, with  
\begin{equation}\label{t1}
    T_{\mu\nu}^{\phi} = -\frac{1}{2}\partial_{\mu}\phi\partial_{\nu}\phi+\frac{1}{4}\partial_{\alpha}\phi \partial^{\alpha}\phi g_{\mu\nu}-\frac{1}{2}\mathcal{V}(\phi)g_{\mu\nu},
\end{equation}
and
\begin{align*}\label{t2}
    T_{\mu\nu}^{LV} &= \frac{1}{2}H^{\alpha\beta}{}{}_{\mu}H_{\nu\alpha\beta} - \frac{1}{12}g_{\mu\nu}H^{\alpha\beta\lambda}H_{\alpha\beta\lambda}\\&-g_{\mu\nu}V + 4B^{\alpha}{}_{\mu}B_{\alpha\nu}V_{X}+ \xi \bigg( \frac{1}{2} B^{\alpha\beta}B^{\lambda\sigma}R_{\alpha\beta\lambda\sigma}g_{\mu\nu} \\&+ \frac{3}{2}B^{\beta\lambda}B^{\alpha}{}_{\mu}R_{\nu\alpha\beta\lambda}  + \frac{3}{2}B^{\beta\lambda}B^{\alpha}{}_{\nu}R_{\mu\alpha\beta\lambda} \\&+\nabla_{\alpha}\nabla_{\beta}(B^{\alpha}{}_{\mu}B_{\nu}{}^{\beta})+\nabla_{\alpha}\nabla_{\beta}(B^{\alpha}{}_{\nu}B_{\mu}{}^{\beta}) \bigg).\numberthis
\end{align*}
On the other hand, by varying the action~\eqref{ac1} with respect to \(\phi\) and \(B_{\mu \nu}\), one obtains, respectively, the equations of motion for the scalar field and for the antisymmetric 2-tensor field, namely 
\begin{align} \label{kkl}
  \Box \phi = &-\frac{\partial \mathcal{V}}{\partial\phi},\\
\label{eomb}\nabla_{\alpha}H^{\alpha\mu\nu} = &\,\,4 V_X B^{\mu\nu}  - \frac{2\xi}{\kappa}B_{\alpha\beta}R^{\alpha\beta\mu\nu}.
\end{align}
where $\Box =\nabla_{\alpha}\nabla^{\alpha} $. We point out that the last equation provides a relation between the LV dynamics and the curvature tensor. As we shall see, in the case of a background field configuration, with an appropriate choice of the VEV of the antisymmetric rank-2 tensor, we can derive important geometric constraints that restrict the geometries that solve the system~\eqref{eq:einstein_KR}.

\section{Lorentz-violating wormholes} \label{seciv}
In this section, we will investigate the effects of the VEV field, i.e., \( \langle B_{\mu\nu} \rangle = b_{\mu \nu}  \), in a static and spherically symmetric spacetime. Moreover, as proposed in Ref.~\cite{Lessa:2019bgi}, it is useful for the purposes of this article to express the VEV as
\begin{equation}
    b_{\mu \nu} = \tilde{E}_{[\mu} v_{\nu]} + \epsilon_{\mu \nu \alpha \beta} v^{\alpha} \tilde{B}^{\beta},
\end{equation}
where the background vectors \( \tilde{E}_{\mu} \) and \( \tilde{B}_{\mu} \) can be interpreted as pseudo-electric and pseudo-magnetic fields, respectively, \( v^{\mu} \) is a timelike 4-vector, and the square brackets in the indexes denote anti-symmetrization, namely $A_{[\mu\nu]}=\tfrac{1}{2}(A_{\mu\nu}-A_{\nu\mu})$. Additionally, the pseudo-fields \( \tilde{E}_{\mu} \) and \( \tilde{B}_{\mu} \) are spacelike, i.e., $
    \tilde{E}_{\mu} v^{\mu} = \tilde{B}_{\mu} v^{\mu} = 0.$

In this work, we consider a pseudo-electric configuration of form~\cite{Lessa:2019bgi}
\begin{equation}
\label{vevansatz}
 \boldsymbol{b}_{2} = - \tilde{E}(x) \ dt \wedge dx,
\end{equation}
i.e., the only nonvanishing terms are $b_{tx}=-b_{xt}= - \tilde{E}(x)$, which can be determined through Eq. (\ref{norm}) for a given geometry. Furthermore, it is straightforward to demonstrate that the strength field $H_{\mu\nu\alpha}$ vanishes for Eq. (\ref{vevansatz})~\cite{Lessa:2019bgi}.

Henceforth, we consider spherically symmetric spacetimes with line element given by
\begin{equation} \label{metric}
   ds^2 = -A(x)dt^2 + \dfrac{dx^2}{A(x)} + r(x)^2 d\Omega^2
\end{equation}
where radial coordinate $ x \in (-\infty,\infty)$. Since we are looking for wormhole solutions, a throat-like structure is present in the spacetime, therefore the areal radius $r(x)$ must possess a local minimum, that is,
\begin{equation}
    r(x_{th})=a,\quad r'(x_{th})=0,\quad r''(x_{th})>0,
\end{equation}
where $a$ and $x_{th}$ are the radius of the wormhole throat and its coordinate location, respectively, and the prime denotes derivatives with respect to $x$,

The gravitational field equations lead to a system of coupled differential equations
\begin{widetext}
\begin{align}
    \label{eq_system_1}&\frac{-1+A(r')^2+r(A'r'+2Ar'')}{r^2} -\frac{l}{2r^2}(2A(r')^2-r^2A''+r(A'r'+2Ar''))-\kappa\left(\frac{A(\phi')^2}{2} -\mathcal{V}\right)-2b^2V_X=0,\\
    \label{eq_system_2}&\frac{-1+rA'r'+A(r')^2}{r^2}-\frac{l}{2r}(A'r'-rA'')+\kappa\left(\frac{A(\phi')^2}{2} +\mathcal{V}\right)-2b^2V_X=0,\\
    \label{eq_system_3}&\frac{2A'r'+rA''+2Ar''}{r}-\frac{l}{2}A''-\kappa\left(\frac{A(\phi')^2}{2} -\mathcal{V}\right)=0,
\end{align}
\end{widetext}
where we have introduced $l = 2 b^2 \xi$. While, the Klein-Gordon Eq.~\eqref{kkl} reads
\begin{equation}
   \bigg[ A\phi''+\left(\frac{2Ar'}{r}+A'\right)\phi' \bigg]  = - \frac{\partial \mathcal{V}}{\partial \phi}.
\end{equation}

It can be observed that by substituting Eqs. (\ref{vevansatz}) and (\ref{metric}) into Eq. (\ref{norm}) yields a constant pseudo-electric field, expressed as
\begin{equation}\label{vv}
    \tilde{E} = \frac{b}{\sqrt{2}}.
\end{equation}
Therefore, the VEV equation of the antisymmetric rank-2 tensor field~\eqref{eomb} reduces to
\begin{equation}
\label{eq:VEV_V_constraint}
2V_X+\frac{\xi}{\kappa}A'' = 0.
\end{equation}
We note that the dynamical Eq.~(\ref{eomb}) becomes a geometric constraint in this configuration. It is important to note that, unlike the non-minimal couplings of background fields to the Ricci tensor -- where the VEV equation results in a differential equation for both the lapse function and the areal radius -- in this case, with only a non-minimal coupling to the Riemann tensor, the constraint imposed by the VEV equation affects solely the lapse function (for a comprehensive study of the emergence and consequences of this type of constraints in Lorentz-violating gravity theories, see Ref.~\cite{Lessa:2025kln}). Such mild constraint is the main reason behind our choice of $S_{LV}$. Since the VEV equation does not impose any constraint on the areal radius, this framework seems to be a good candidate for the search of wormhole configurations. We point out that, by vanishing the LV parameter, the anisotropy introduced by the LV disappear and no constraints is imposed on the metric field.

Here we focus on two configurations of potential $V$ that trigger spontaneous breaking of Lorentz symmetry, namely $V^Q(X)= \frac{1}{2}\zeta X^2$ and $V^L(X)=\lambda X$. 
\subsection{Quadratic potential}
Let us first consider the quadratic potential case (\ref{quadr}). In this scenario, both $\langle V^Q\rangle$ and $\langle V^Q_X\rangle$ vanish. Therefore, Eq.~\eqref{eq:VEV_V_constraint} reduces to
\begin{equation}
A''\equiv 0. 
\end{equation}
Hence, the form of the lapse function is constrained, and for the quadratic potential the most general form is given by
\begin{equation}
    A(x) = a_1 + a_2 x,
\end{equation}
where $a_1$ and $a_2$ are two integration constants. 

\subsubsection{Constant $A(x)$: No Killing horizons}\label{sec:noKilling}
Before we delve into the most general lapse function allowed, it is worth considering lapse functions with a constant profile $A(x)=a_1$. In such case, no Killing horizon is present in the spacetime. A particular interesting case is $A(x)=1$.

By subtracting Eq.(\ref{eq_system_1}) from Eq.(\ref{eq_system_3}), we obtain the following equation, which depends only on the areal function:
\begin{equation}
\label{eq_radial_ellis_like} 1+(l-1)((r')^2+rr'')=0.
\end{equation}
One can solve this differential equation with initial conditions $r(0)=a$ and $r'(0)=0$, finding that
\begin{equation}
    \label{eq:r_solution_ellis}r(x)=\sqrt{a^2+\frac{x^2}{1-l}},
\end{equation}
where we have put the throat location at $x=0$. From Eq.~\eqref{eq_radial_ellis_like} it follows that
\begin{equation}
    r''(0)=\frac{1}{(1-l)a},
\end{equation}
and therefore, to ensure the existence of a throat, the Lorentz violating parameter is constrained by $l<1$. 

\begin{figure}[!h]
    \centering
    \includegraphics[width=\linewidth]{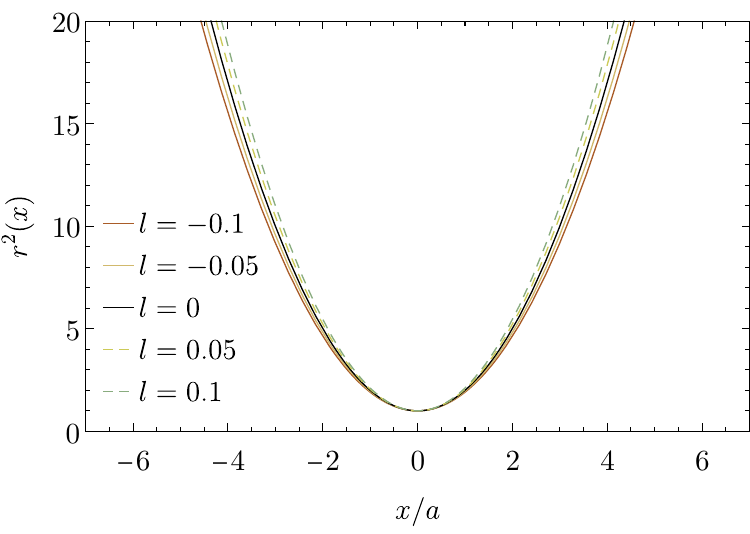}
    \caption{Areal radius squared for LVEB wormholes.}
    \label{fig:areal_radius_LVEB}
\end{figure}

The resulting spacetime is described by
\begin{equation}
  \label{eq:ellis_wormhole}  ds^2 = -dt^2+dx^2+\left(a^2+\frac{x^2}{1-l}\right)d\Omega^2.
\end{equation}
We notice that in the limit $l \to 0$, the EB solution is recovered and the areal radius becomes $r(x)=\sqrt{a^2+x^2}$. Due to it, henceforth we call the solution~\eqref{eq:ellis_wormhole} Lorentz-violating EB (LVEB) wormhole. We plot in Fig.~\ref{fig:areal_radius_LVEB} the areal radius squared, $r^2(x)$, for some LV parameters, noticing that for smaller values of $l$, $r^2(x)$ is a wider parabola.

\begin{figure}
    \centering
    \includegraphics[width=\columnwidth]{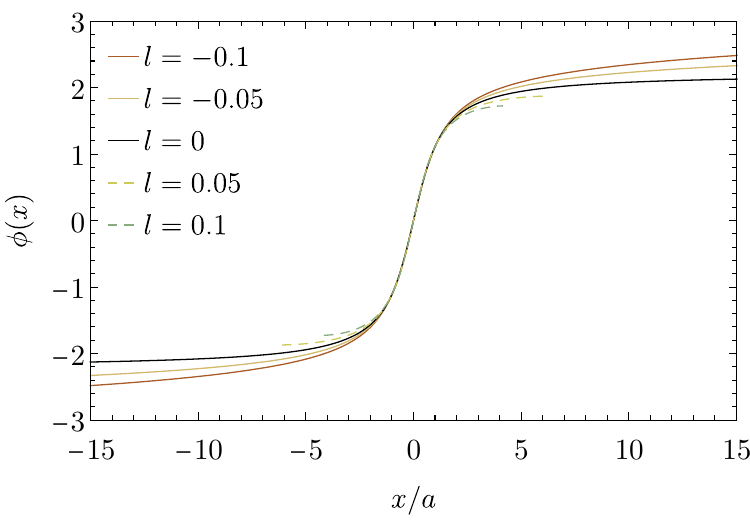}
    \caption{Scalar field solution of some LVEB wormholes. In order to better visualize the transition between purely real and complex scalar fields for $l>0$, we exhibit only the field where $\text{Im}[\phi]=0$.}
    \label{fig:phi_LVEB}
\end{figure}

By substituting the solution (\ref{eq:r_solution_ellis}) into Eqs. (\ref{eq_system_1}) and (\ref{eq_system_2}), and then subtracting them, we find that the scalar field satisfies
\begin{equation}
\label{eq:eq_scalar_field_ellis} \kappa^2(\phi')^2=\frac{a^2(l-2)(l-1)-lx^2}{(x^2-a^2(l-1))^2}.
\end{equation}
By considering the initial condition $\phi(0)=0$, the general solution of the above differential equation reads
\begin{equation}
\label{eq:phi_sol_ellis}    \phi = \phi_0+\sqrt{\frac{2}{\kappa}}\arctan\left(\frac{\sqrt{2}x}{\bar{x}}\right)-\sqrt{\frac{-l}{\kappa}}\log\left(2(\bar{x}-x\sqrt{-l})\right),
\end{equation}
where $\phi_0=-\sqrt{-l/\kappa}\log(2a\sqrt{(l-2)(l-1)})$ and $\bar{x}=\sqrt{a^2(l-2)(l-1)-lx^2}$. It follows that, in order to have a real scalar field supporting the LVEB wormhole, the LV parameter $l\leq 0$. This fact can also be anticipated from the inspection of Eq.~\eqref{eq:eq_scalar_field_ellis}, where one notices that, the right hand side of it is non-negative for $l\leq 0$ regardless the value of $x$. Hence the derivative of the scalar field is a real-valued function, so the scalar field is real. In the limit $l\to0$ the scalar field reduces to the phantom scalar field that supports the EB wormhole, namely $\phi(x)=\sqrt{2/\kappa}\arctan(x/a)$.
In Fig.~\ref{fig:phi_LVEB} we show the scalar fields that supports some LVEB wormholes. For $l>0$ we show only the scalar field where $\text{Im}[\phi]=0$.

Different from GR where free phantom scalar fields can support wormholes, in the Lorentz-violating scenario considered here, phantom wormholes require a self-interaction term in the scalar field Lagrangian. By substituting Eq.~\eqref{eq:r_solution_ellis} and Eq.~\eqref{eq:phi_sol_ellis} into Eq.~\eqref{eq_system_2}, we find that
\begin{equation}
    \mathcal{V} =\frac{l}{2 \kappa  \left(a^2 (l-1)-x^2\right)}.
\end{equation}
In the limit \( l \to 0 \), this potential vanishes, what is in agreement with the line element reducing to the EB one. 

It is worth noting that, the same line element can be obtained through the choice of different constant lapse function $A(x)=1/(1-l)$. Under this assumption, the subtraction of Eq.~\eqref{eq_system_1} and Eq.~\eqref{eq_system_3} results in
\begin{equation}
    1-(r')^2-rr''=0.
\end{equation}
By considering the initial conditions that correspond to a throat located at $x=0$, the solution of the above equation is the standard EB areal radius, that is, $r(x)=\sqrt{a^2+x^2}$. The resulting line element is therefore written as
\begin{equation}
\label{eq:line_element_asymp}    ds^2=-\frac{1}{1-l}dt^2+(1-l)dx^2+(x^2+a^2)d\Omega^2.
\end{equation}
By performing a coordinate transformation, namely $t\to\tilde{t}/\sqrt{1-l}$ and $x\to\tilde{x}\sqrt{1-l}$, and dropping the tilde we recover line element~\eqref{eq:ellis_wormhole}. Writing the line element of the EB-like wormhole as~\eqref{eq:line_element_asymp} reveals that such phantom wormhole has the same asymptotic as black hole solutions found in Lorentz-violating scenarios, either considering an antisymmetric rank-2 tensor field field non-minimally coupled to gravity~\cite{Yang:2023wtu,duan2024electrically} or the bumblebee field coupled to gravity~\cite{Casana:2017jkc}. 

\subsubsection{Affine $A(x)$: Killing horizons}\label{sec:affine_A}
Now, let us turn our attention to lapse functions with linear contributions, that is, $A(x)=a_1+a_2 x$. This type of contribution appears, for instance, in Kiselev solutions with appropriate equation of state~\cite{kiselevQuintessenceBlackHoles2003}, vacuum solutions of conformal Weyl gravity~\cite{mannheimExactVacuumSolution1989} and Cotton gravity~\cite{gogberashviliGeneralSphericallySymmetric2024}, and in model for gravity at large distances~\cite{grumillerModelGravityLarge2010}. The vanishing of $A(x)$ represents a Killing horizon, and for such form it is located at $x_h=-a_1/a_2$. Therefore, if a throat is present, the wormhole is asymmetric once there is a horizon in just one side. Let us look for such asymmetric wormholes. For simplicity, let us consider a lapse function of the form $A(x)=1+\chi x$, where $\chi$ can be seen as a Rindler-type acceleration.

By subtracting Eq.~\eqref{eq_system_1} from Eq.~\eqref{eq_system_3}, we obtain that the areal radius satisfies
\begin{equation}
    1+\frac{1}{2}l\chi rr'+(1+\chi x)(l-1)((r')^2+rr'')=0.
\end{equation}
By choosing the corresponding initial conditions of a throat located at $x=0$, it follows that, for $l\neq0$, 
\begin{equation}
\label{eq:radius_quintessence}    r(x) = \sqrt{a^2+\frac{2(2N_1-1)(-1-N_1 x \chi +(1+x\chi)^{N_1})}{(N_1-1)N_1\chi^2}},
\end{equation}
where $N_1=(l-2)/(2l-2)$. For the sake of comparison, it worth considering the the limit $l\to0$ ($N_1\to 1$), where the areal radius goes to the solution of the Eq.~\eqref{eq_system_1} with $l=0$, namely  
\begin{equation}\label{l0}
    r(x)=\sqrt{a^2+\frac{2(1+\chi x)\log(1+\chi x)-2\chi x}{\chi^2}}.
\end{equation}

We point out that in the limit $\chi\to 0$, Eq.~\eqref{eq:radius_quintessence} reduces to Eq.~\eqref{eq:r_solution_ellis}. One notices that, at $x=0$, $r(0)=a$, $r'(0)=0$ and $r''(0) = 1/a(1-l)$, hence $l<1$ in order to guarantee a local minimum, therefore $N_1\in(1/2,\infty)$. One verifies that $r^2(x)$ is symmetric with respect to the throat only $N_1=2$, that is $l=2/3$. Any other value of $l$ turns the areal radius asymmetric. If $N_1\neq 2$, $(1+x\chi)^{N_1}$ becomes complex whenever $(1+x\chi)$ is negative. This happens in the region beyond the Killing horizon and therefore in a causally disconnected region. Due to the Rindler-type acceleration in the resulting wormhole spacetime, namely
\begin{align*}
    \label{eq:ALV_wormhole}ds^2 =& -(1+\chi x)dt^2+\frac{1}{1+\chi x}dx^2+\\&+\left[a^2+\frac{\beta(-1-N_1 x \chi +(1+x\chi)^{N_1})}{\chi^2}\right]d\Omega^2\numberthis,
\end{align*}
where $\beta=2(2N_1-1)/(N_1-1)N_1$. There is a sort of Rindler-like character in this spacetime that shows up by performing the following redefinitions
\begin{equation}
    x=\frac{(1-y^2\alpha^2)}{2\alpha},\quad \chi=2\alpha,
\end{equation}
such that the line element~\eqref{eq:ALV_wormhole} becomes
\begin{align*}
    \label{eq:mod_rindler}ds^2=&-\alpha^2 y^2dt^2+dy^2\\&+\left[a^2+\beta\frac{(N_1(1-\alpha^2y^2)+(\alpha^2y^2)^{N_1}-1)}{\alpha^2} \right]d\Omega^2\numberthis.
\end{align*}
With this redefinition the throat is located at $y=1/\alpha$ and the horizon at $y=0$. Despite the spherical sector, we restore the same behavior of 2D Rindler spacetime with acceleration $\alpha$. The line element~\eqref{eq:ALV_wormhole} can then be understood as a wormhole spacetime with a Rindler-type acceleration with a modified spherical sector that depends on the acceleration and on the LV parameter, and henceforth we call this solution accelerated Lorentz-violating wormhole, for short ALV wormhole. One can verify that in the limit $\chi\to 0$, the line element~\eqref{eq:ellis_wormhole} is restored.

\begin{figure}[!htb]
    \centering
    \includegraphics[width=\columnwidth]{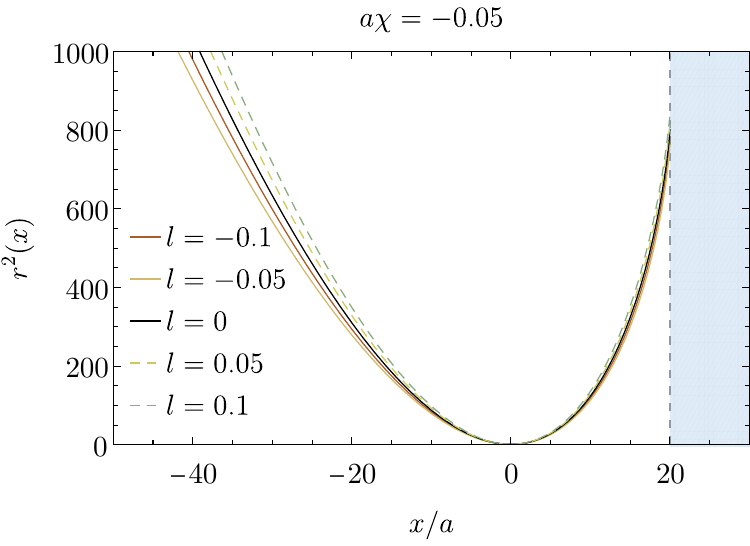}  
    \caption{Areal radius squared for some ALV wormhole configurations. The Blue shaded region corresponds to the region beyond the Killing horizon.}
    \label{fig:areal_radius_ALV}
\end{figure}

In Fig.~\ref{fig:areal_radius_ALV} we show the areal radius squared for some ALV wormholes, where we notice that for smaller $l$ parameters, $r^2(x)$ has a slower growth far from the (Rindler) horizon. As one approaches the horizon, the parameter $\chi$ dominates the behavior of $r^2(x)$, and it approaches $x=-1/\chi$ with a similar rate, regardless the value of $l$. 

In order to determine the type of scalar fields that supports this asymmetric wormhole we substitute Eq.~\eqref{eq:radius_quintessence} into Eqs.~\eqref{eq_system_1} and~\eqref{eq_system_2} and subtract them obtaining the following differential equation
\begin{equation}
\label{eq:scalar_field_A_linear}    \kappa(\phi')^2=\frac{2 N_1^2 \chi ^2 \, \mathcal{A}}{\mathcal{B}^2},
\end{equation}
where
\begin{align*}
\mathcal{A} = &(\chi x + 1)^{N_1}(N_1 T_1 + 2 \chi x) + (1 - 2 N_1) (\chi x + 1)^{2 N_1} \\&- (1 - 2 N_1)^2 (\chi x + 1)^2, \\
\mathcal{B} =& (\chi x + 1) \big[ N_1 T_2 + 2 ((\chi x + 1)^{N_1} - 1)) \big],\\
T_1 =& N_1 \big( N_1 \chi (a^2 \chi - 4x) - 2 a^2 \chi^2 + 14 \chi x + 4 \big) \\&+ a^2 \chi^2 - 10 \chi x - 2,
\\
T_2 =& N_1 \chi (4x - a^2 \chi) + a^2 \chi^2 - 4 (\chi x + 1)^{N_1} - 2 \chi x + 4.
\end{align*}
Hence, the scalar field can be formally written as 
\begin{equation}
    \phi_{\pm}(x)= \pm N_1\chi\sqrt{\frac{2}{\kappa}}\int_{0}^x \frac{\sqrt{\mathcal{A}}}{\mathcal{B}}dx,
\end{equation}
where we have chosen the scalar field at the throat to vanish. Again, different from GR, a free scalar field cannot close the system, and one needs a self-interacting potential, $\mathcal{V}(\phi)$. To find the explicit dependence on $\phi$ of this potential is complicated due to the complexity of the above integral, actually it is impracticable to look for an analytical expression for the scalar field for arbitrary parameters. Fortunately, by manipulating the differential equations, it is possible to find an expression for the potential as a function only of the areal radius and its derivatives, namely
\begin{equation}
    \mathcal{V} =\frac{2+(l-2)(1+\chi x)r'}{2\kappa r^2}+\frac{(l-2)(\chi r' + (1+\chi x)r'')}{2\kappa r}.
\end{equation}

\begin{figure}[!h]
    \centering
    \includegraphics[width=\columnwidth]{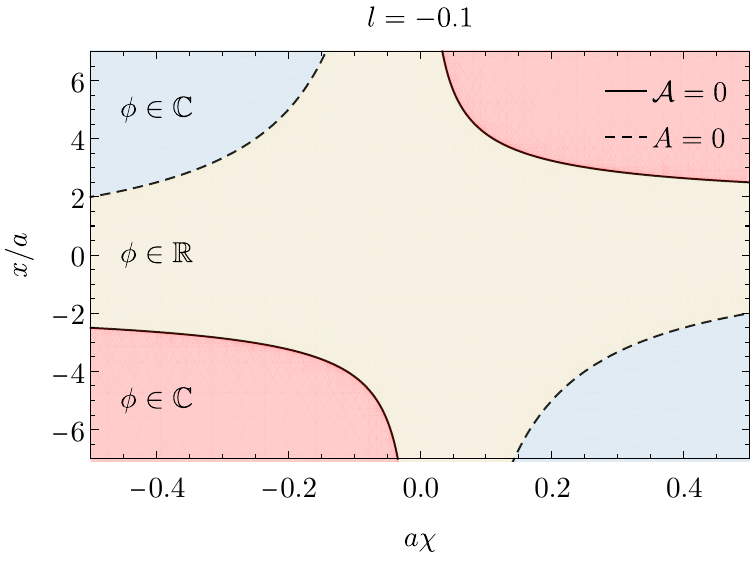}
    \includegraphics[width=\columnwidth]{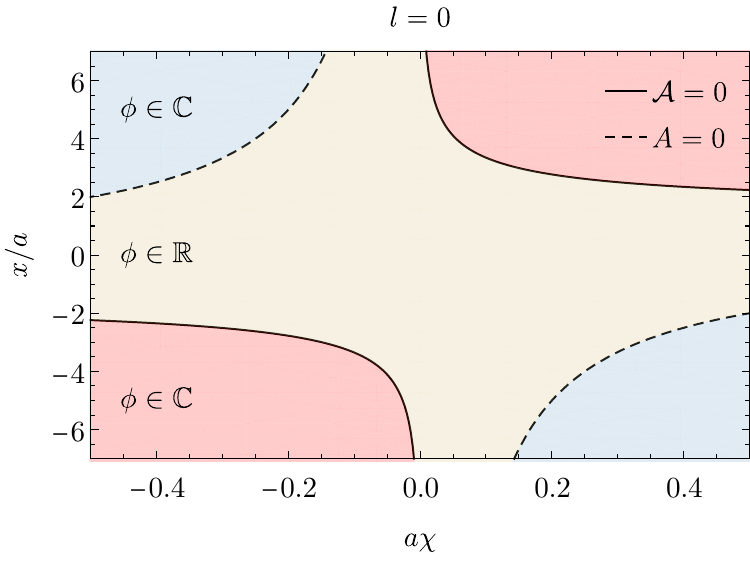}
    \includegraphics[width=\columnwidth]{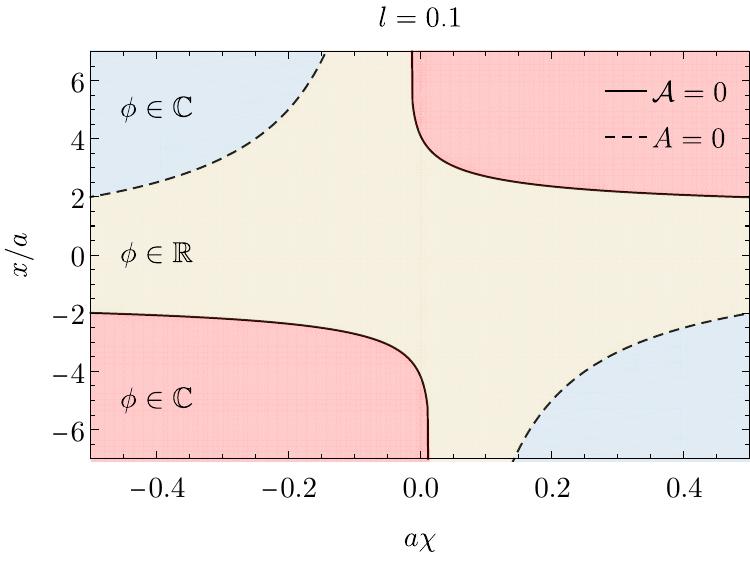}
    \caption{The positiveness of ${\cal A}$ in the $(x,\chi)$ space. The centered yellow region represents where ${\cal A}>0$. The corner blue regions delimited by the dashed line represents the region inside the Killing horizon. The corner red regions delimited by the solid lines represent where ${\cal A}<0$. The dashed and solid line are, respectively, $A=0$ and $\cal A$.}
    \label{fig:RealOrComplex}
\end{figure}

In order to bypass the analytical treatment of the scalar field equation, let us perform a numerical analysis of the scalar field solutions.
But first, we notice that the right hand side of Eq.~\eqref{eq:scalar_field_A_linear} depends on $(1+x\chi)^{N_1}$, therefore in the region beyond the Killing horizon the scalar field is unavoidably complex. Moreover, if the right hand side of Eq.~\eqref{eq:scalar_field_A_linear} is negative, that is $\mathcal{A}<0$, for some value of $x$, the scalar field also becomes complex. It is important to point out that it happens regardless the value of $l$, what is persistent even in the absence of the Lorentz violation. By considering $l=0$, $\mathcal{A}$ becomes
\begin{align*}
    \mathcal{A}=&-((1+\chi x)\log(1+\chi x)-2)(1+\chi x)\log(1+\chi x)\\
    &+\chi(a^2\chi-2x).\numberthis
\end{align*}
One can check that, for arbitrary non-zero values of $\chi$, $\cal A$ becomes negative from some radial coordinate. We show this behavior in the middle panel of Fig.~\ref{fig:RealOrComplex}, where we exhibit for different values of the Rindler-type acceleration, where $\cal A$ is positive (yellow region) or negative (red region). We also exhibit using a blue color the region inside the Killing horizon. As we can see, as we increase the value of $\chi$, the interval where $\cal A$ is positive diminishes. Qualitatively, this same behavior is present in the Lorentz violating case, $l\neq 0$. In the top and bottom panels of Fig.~\ref{fig:RealOrComplex} we exhibit where $\cal A$ is positive or negative, and consequently where the solution is supported by a real or complex phantom scalar field, respectively. As we can see, for $l<0$, the interval where $\cal A$ is positive is wider than for $l>0$. However, unavoidably $\cal A$ becomes negative, and therefore $\phi$ turns into a complex-valued function. 

\begin{figure}[!h]
    \centering
    \includegraphics[width=\columnwidth]{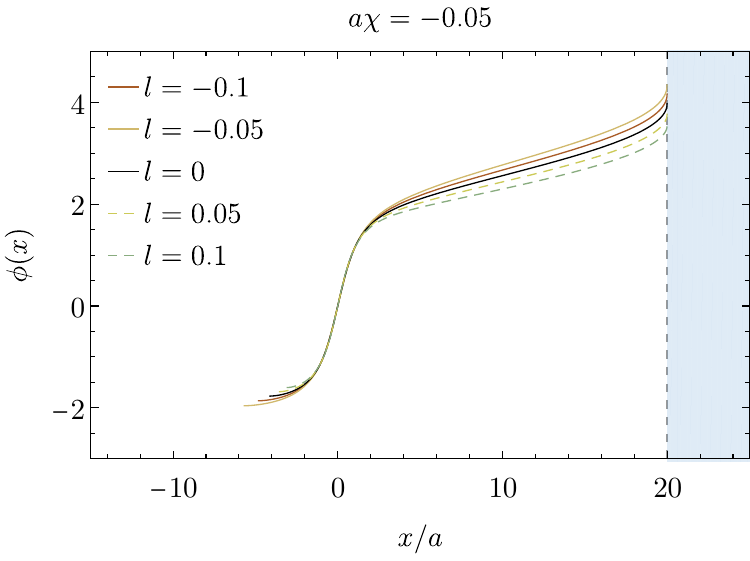} \includegraphics[width=\columnwidth]{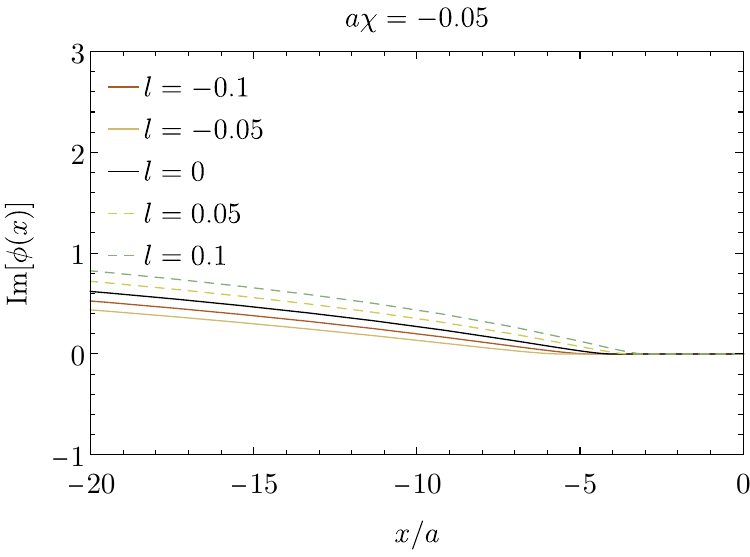}
    \caption{Top panel: Scalar field of some ALV wormholes. In order to better visualize the transition between purely real and complex scalar fields we exhibit only the field where $\text{Im}[\phi]=0$. The blue shaded region represents the region beyond the Killing horizon. Bottom panel: Imaginary part of the scalar field.}
    \label{fig:phi_rindler}
\end{figure}

We show in Fig.~\ref{fig:phi_rindler} the scalar field for some values of $l$. In the top panel of it, we show the purely real part of $\phi$ (where $\text{Im}[\phi]=0$), where we can see that as we decrease $l$, wider is the interval where $\phi$ is real. In the bottom panel of Fig.~\ref{fig:phi_rindler} we show the behavior of the negative part of the scalar field, where we notice that, it boundlessly grows. Whether this transition from real to complex scalar fields implies instabilities in the spacetime will not be addressed here. We remark that a similar scalar field transition happens in the context of low-energy string theory~\cite{horowitz1992dark} and Horndeski theory~\cite{rinaldi2012black}.

An important point to investigate is if it is possible to generate wormhole solutions with Rindler-type acceleration in the context of Lorentz violating gravity supported by an everywhere real (phantom) scalar fields. In Ref.~\cite{anabalon2014asymptotically} they circumvent this issue in Horndeski theory by including a cosmological constant, allowing them to find asymptotically AdS black holes with a real scalar field outside their horizons. Remarkably, in Lorentz violating contexts there is a natural and straightforward way to include an effective cosmological constant, that is by considering a linear potential, $V(X)=\lambda X$, instead of a quadratic one. Let us look for the solutions generated in these scenarios and probe if they can be supported by real scalar fields.

\subsection{Linear potential}
Let us turn our attention to the linear potential case (\ref{line}). In this scenario, while $\langle V^L \rangle$ vanishes, $\langle  V^L_X\rangle$ does not vanish, conversely, it is $\langle V^{L}_X\rangle=\lambda$. Hence, Eq.~\eqref{eq:VEV_V_constraint} yields to
\begin{equation}
    2\kappa\lambda+\xi A''=0.
\end{equation}
Therefore, the lapse function is constrained, similarly to the quadratic case, and its most general form is
\begin{equation}
    A(x)=\tilde{a}_1+\tilde{a}_2 x-\frac{\kappa\lambda}{\xi}x^2,
\end{equation}
where $\tilde{a}_1$ and $\tilde{a}_2$ are two integration constants. As we can see, the lapse function has a quadratic dependence on the $x$, linear on the Lagrange multiplier $\lambda$. 

In general, quadratic lapse functions appears as a characteristic of solutions endowed with a cosmological constant $\Lambda$, such as solutions with de-Sitter (dS) and Anti-de-Sitter (AdS) asymptotic. By calling $\lambda=\xi \Lambda_{e}/3\kappa$, the lapse function reduces to
\begin{equation}
    A(x)=\tilde{a}_1+\tilde{a}_2x-\frac{\Lambda_e}{3}x^2,
\end{equation}
where $\Lambda_e$ can be interpreted as an effective cosmological constant. Since the location of the Killing horizons can be found as the roots of $A(x)=0$, if $r(x)$ has a local minimum (a throat is present), the resulting spacetime can have two distinct cosmological horizons, located, respectively, at
\begin{equation}
\label{eq:cosmological_horizons}    x_{\pm} = \frac{3\tilde{a}_2\pm\sqrt{9\tilde{a}_2^2+12\tilde{a}_1\Lambda_e}}{2\Lambda_e}.
\end{equation}
Here, let us give attention to the case $A(x)=1+\chi x - \Lambda_ex^2/3$, and look for wormhole solutions endowed with a cosmological constant and a Rindler-type acceleration.

By subtracting Eq. (\ref{eq_system_1}) from Eq. (\ref{eq_system_3}), we obtain the following equation
\begin{align*}
\label{eq:cosmological_ellis1}    &-6+(2-3l)\Lambda_er^2-2(l-1)(3+3\chi x -\Lambda_e x^2)(r')^2
    \\&+r[(2\Lambda_e x - 3\chi)l r'-2(l-1)(3+3\chi x -\Lambda_e x^2)r'']=0,\numberthis
\end{align*}
while by subtracting Eq. (\ref{eq_system_1}) from Eq. (\ref{eq_system_2}) yields
\begin{equation}
\label{eq:cosmological_ellis2}    \kappa r^2(\phi')^2+l(r')^2+(l-2)rr''=0.
\end{equation}
In order to close the system, a self-interaction potential is required. By manipulating the differential equations, it is possible to find an expression for the potential as a function only of the areal radius and its derivatives, namely
\begin{align*}
    \mathcal{V} =&\,\frac{6+r^2(6l\Lambda_e+\kappa(\Lambda_e x^2-3\chi x-3)(\phi')^2)}{6\kappa r^2}\\&+\frac{(l-2)(3\chi-2\Lambda_e x)rr'}{6\kappa r^2}+\frac{2(\Lambda_e x^2-3\chi x -3)(r')^2}{6\kappa r^2}.\numberthis
\end{align*}

Differently from the two previous systems discussed in the last subsection, the radial equation~\eqref{eq:cosmological_ellis1} cannot be solved analytically for arbitrary value of $\Lambda_e$. Therefore, we investigate the solutions of this system only numerically by using a Runge-Kutta integrator. We recall that, in order to have a throat at $x=0$, $r'(0)=0$ and
\begin{equation}
    r''(0) = \frac{6-a^2\Lambda_e(-2+ 7l)}{6a(1-l)}>0.
\end{equation}
Contrary to the quadratic potential case, when an effective cosmological constant is present, one can also find throats at $x=0$ for $l>1$ if $\Lambda_e<-6/a^2(-2+7l)$. Even though such configurations are allowed, here we focus in small Lorentz violation parameters, therefore $|l|< 1$.

By rearranging Eq.~\eqref{eq:cosmological_ellis2}, one obtains that the scalar field $\phi$ is real only if
\begin{equation}
    \label{eq:cosmological_ellis3} -l(r')^2-(l-2)rr'' \geq 0.
\end{equation}
Since the areal radius is supposed to be a non-negative function and we are assuming $|l|< 1$, whenever 
\begin{equation}
    \label{eq:cond_field} r''\geq \frac{-l (r')^2}{(l-2)r},
\end{equation}
$(\phi')^2$ is positive and therefore $\phi\in \mathbb{R}$.

Let us split the analysis in two distinct groups of solutions, namely the ones with positive $\Lambda_e$ and the ones with negative $\Lambda_e$. 

\subsubsection{Negative $\Lambda_e$}
By considering negative values of $\Lambda_e$, 
 $x_\pm$ can take either reals or complex values
 [see Eq.~\eqref{eq:cosmological_horizons}]. In particular, if $|\chi|<2\sqrt{-\Lambda_e/3}$, $x_\pm\in \mathbb{C}$ and no Killing horizon is present in the spacetime. In this case, if wormhole solutions arise, they have two Anti-de-Sitter-like asymptotic, but as we shall see, with an angular sector that deviates from a parabola.

\begin{figure*}
\includegraphics[width=\columnwidth]{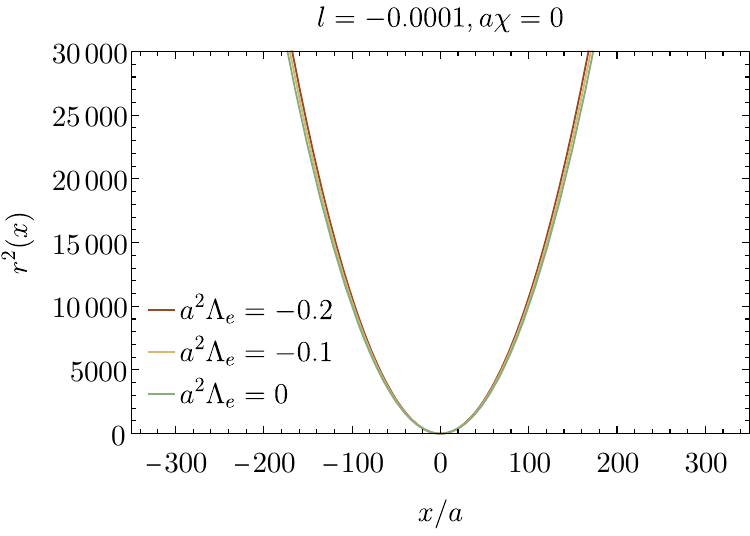}
\includegraphics[width=\columnwidth]{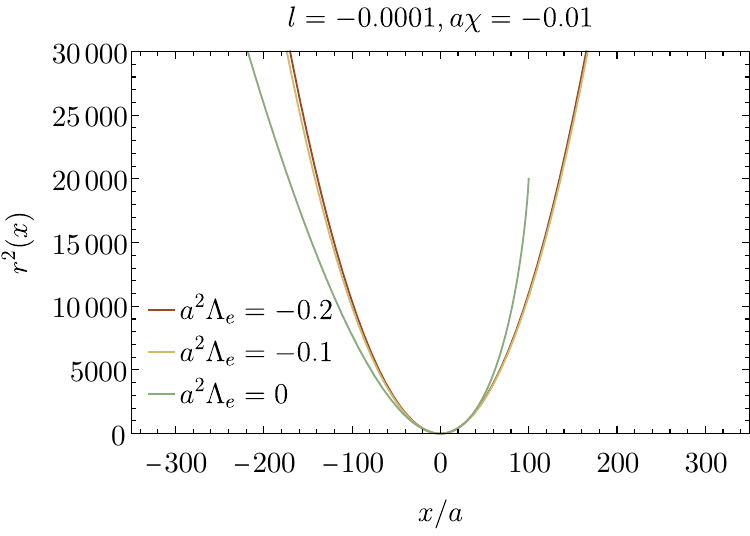}
\caption{Areal radius squared for LV wormholes with negative cosmological constant.}
\label{fig:ads_areal_radius}
\end{figure*}

\begin{figure*}
    \centering
    \includegraphics[width=\columnwidth]{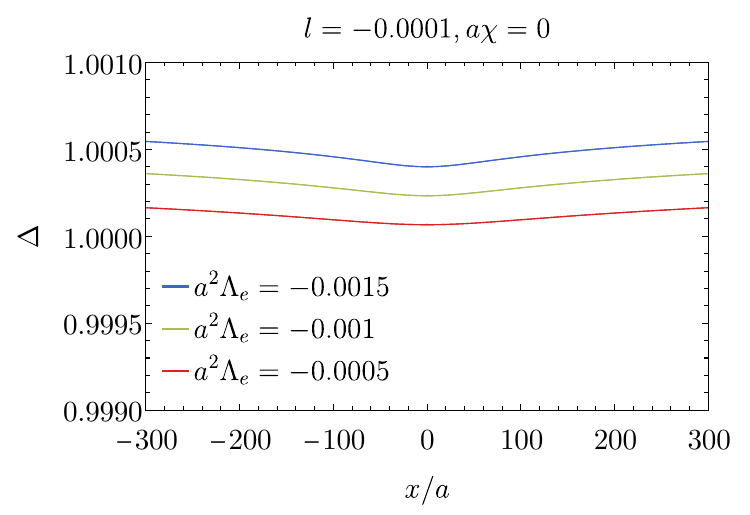}
    \includegraphics[width=\columnwidth]{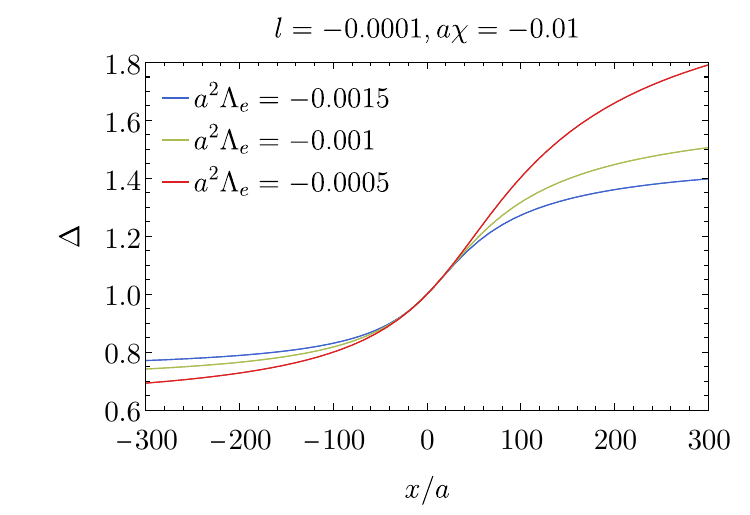}
    \caption{The deviation from a parabola of the areal radius squared of some LV wormholes with negative cosmological constant.}
    \label{Fig:Delta_Lambda_neg}
\end{figure*}

 In order to obtain wormhole solutions, we numerically integrate Eq.~\eqref{eq:cosmological_ellis1} subjected to the initial conditions $r(0)=a$ and $r'(0)=0$. We show in Fig.~\ref{fig:ads_areal_radius} a selection of areal radius squared for some negative values of $\Lambda_e$. We notice that, in the presence of an effective cosmological constant, the areal radius deviates from a parabola either for vanishing and non-vanishing values of $\chi$. In order to visualize this deviation, we compute the quantity
 \begin{equation}
     \Delta = \frac{r^2(x)-a^2}{x^2}.
 \end{equation}
 If $r^2(x)$ had a parabolic behavior, then such quantity would be a constant as in the LVEB wormhole case [see Sec.~\ref{sec:noKilling}], namely
 \begin{equation}
     \Delta_{LVEB} = \frac{a^2+\dfrac{x^2}{1-l}-a^2}{x^2}=\frac{1}{1-l}.
 \end{equation}
We show in Fig.~\ref{Fig:Delta_Lambda_neg} the function $\Delta$ for some negative values of the effective cosmological constant. As we can see, it is not a constant, and therefore the areal radius, in neither side of the wormhole, has a parabolic behavior.

 We recall that, the wormhole is symmetric if $\chi=0$ and asymmetric if $\chi\neq 0$. As already pointed out, if $|\chi|<2\sqrt{-\Lambda_e/3}$ neither side of the wormhole has a horizon. In both sides, the areal radius has a faster growth as $\Lambda_e$ decreases. In particular, if $\chi<0$ ($\chi<0)$, the growth is faster for $x<0$ ($x>0$).

\begin{figure*}
\includegraphics[width=\columnwidth]{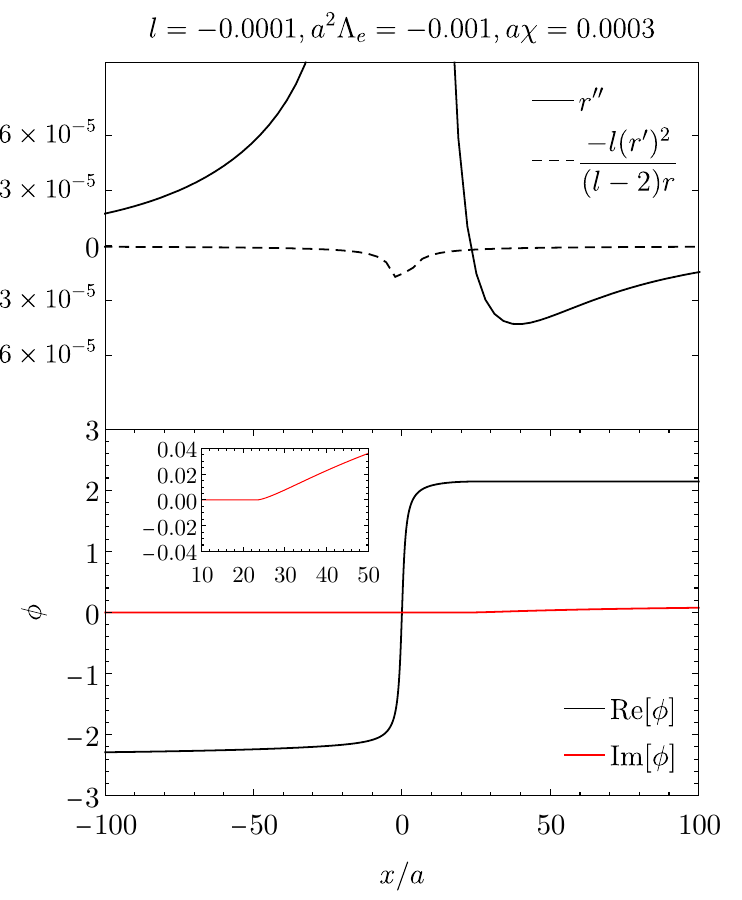}
\includegraphics[width=\columnwidth]{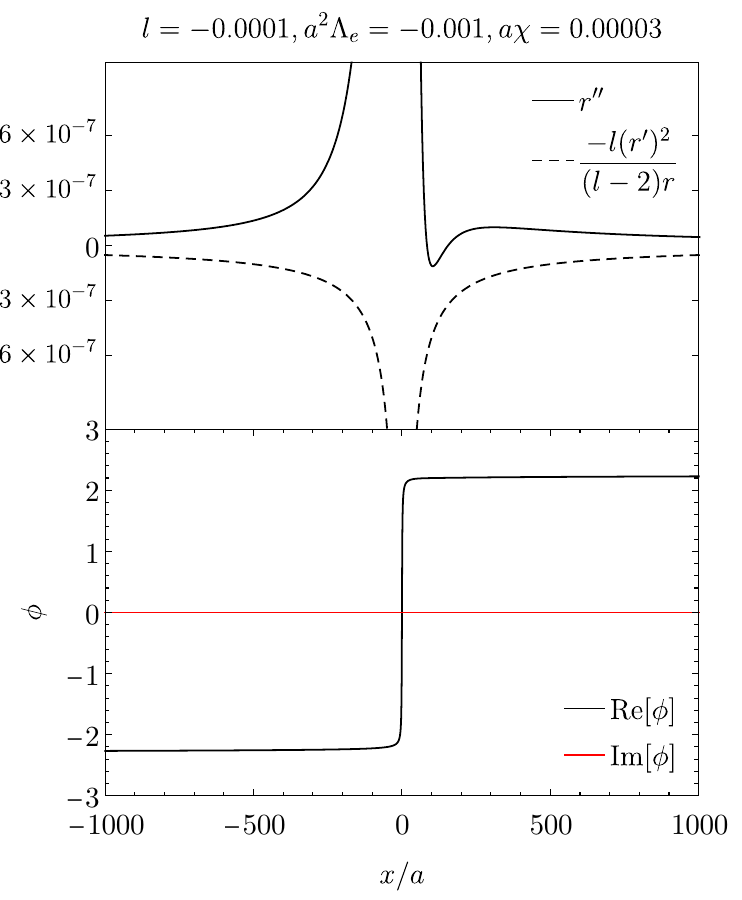}
\caption{Condition for the realness of the scalar field of LV wormholes with negative cosmological constant. In the left panel we show a scalar field that becomes complex from a value of the radial coordinate and in the right panel we show a scalar field that is purely real regardless the value of $x$.}
\label{fig:ads_scalar_field}
\end{figure*}

 It is worth noting that, different from the wormhole solution discussed in Sec.~\ref{sec:affine_A}, the presence of the negative effective cosmological constant enables wormholes to be supported by real scalar fields. Once the (numerical) areal radius is obtained, one can verify the condition~\eqref{eq:cond_field}. We show this in Fig.~\ref{fig:ads_scalar_field} for two distinct scenarios. In the left panel, the condition~\eqref{eq:cond_field} is not fulfilled, therefore the scalar field turns into a complex-valued function starting from a value of the radial coordinate $x$. On the other hand, in the right panel, the condition~\eqref{eq:cond_field} is fulfilled, and the scalar field is a real-valued function in the whole domain of $x$. 
 
\subsubsection{Positive $\Lambda_e$}
Considering positive values of $\Lambda_e$, two Killing horizons unavoidably appears in the spacetime. The location of these cosmological horizons are 
\begin{equation}
    x_\pm = \frac{3\chi\pm\sqrt{12\Lambda_e+9\chi^2}}{2\Lambda_e}.
\end{equation}
Therefore, wormholes that arise in such context have two de-Sitter-like asymptotic, nonetheless, as we shall see, with an angular sector that deviates from a parabola.  

\begin{figure*}
\includegraphics[width=\columnwidth]{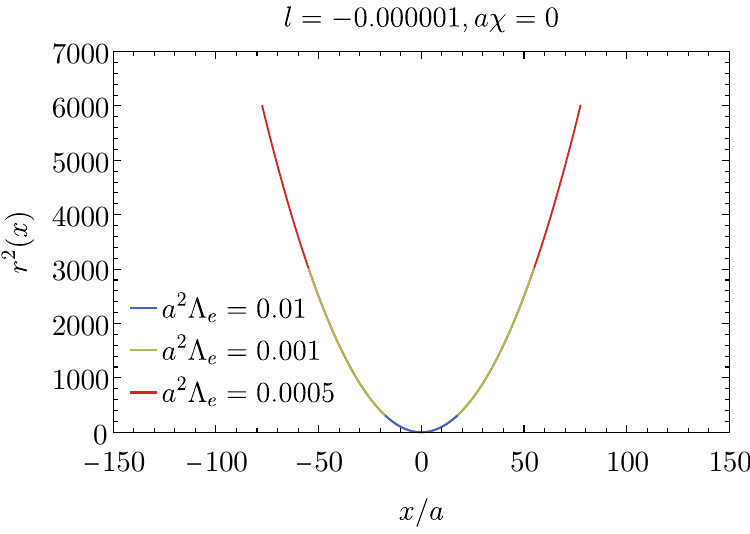}
\includegraphics[width=\columnwidth]{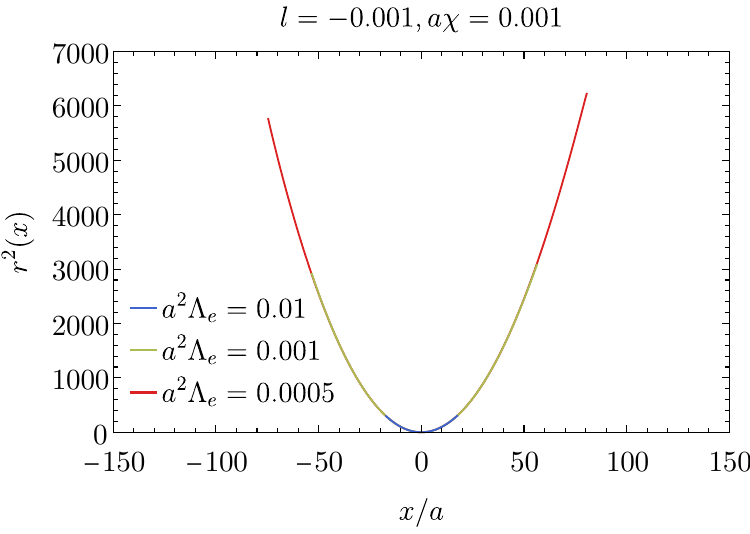}
\caption{Areal radius squared for LV wormholes with positive cosmological constant.}
\label{fig:ds_areal_radius}
\end{figure*}

\begin{figure*}
    \centering
    \includegraphics[width=\columnwidth]{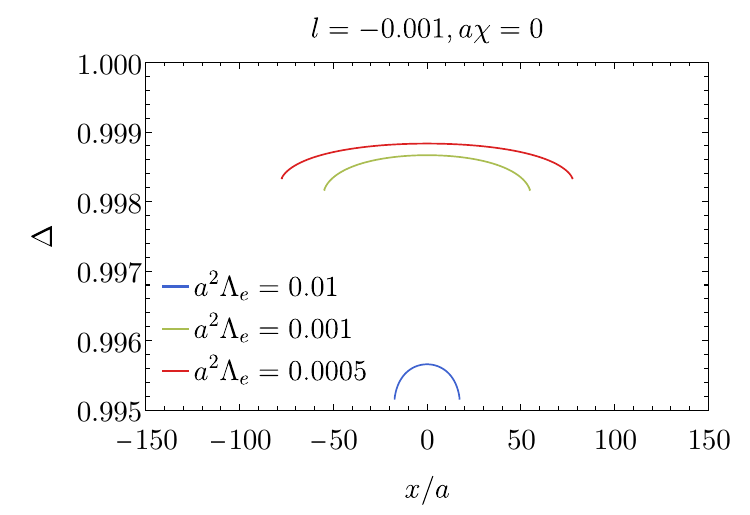}
    \includegraphics[width=\columnwidth]{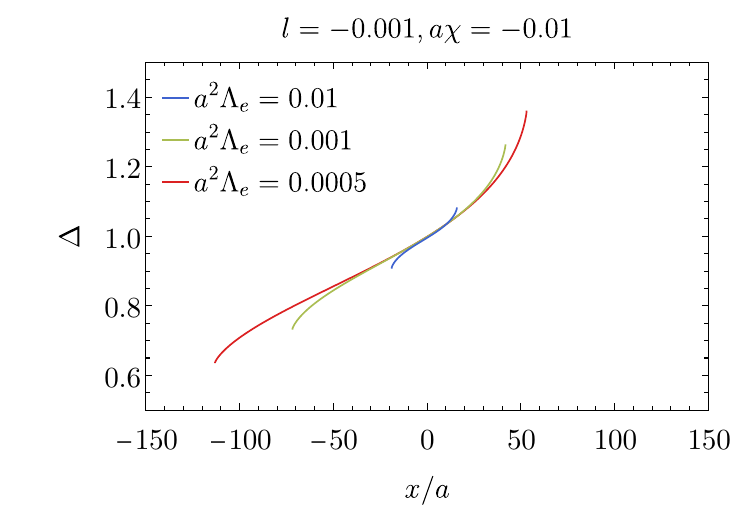}
    \caption{The deviation from a parabola of the areal radius squared of some LV wormholes with positive cosmological constant.}
    \label{Fig:Delta_Lambda_pos}
\end{figure*}

 In order to obtain wormhole solutions, we numerically integrate Eq.~\eqref{eq:cosmological_ellis1} considering the the initial conditions $r(0)=a$ and $r'(0)=0$. We show in Fig.~\ref{fig:ds_areal_radius} a selection of areal radius squared for some positive values of $\Lambda_e$. We point out that, in the presence of a positive effective cosmological constant, the areal radius deviates from a parabola either for vanishing and non-vanishing values of $\chi$. To see this, we exhibit in Fig.~\ref{Fig:Delta_Lambda_pos} the function $\Delta$ for different values of $\Lambda_e$. We notice that $\Delta$ is not a constant, and therefore $r^2(x)$ has a non-parabolic behavior between the two cosmological horizons.
 
\begin{figure*}[!h]
\includegraphics[width=\columnwidth]{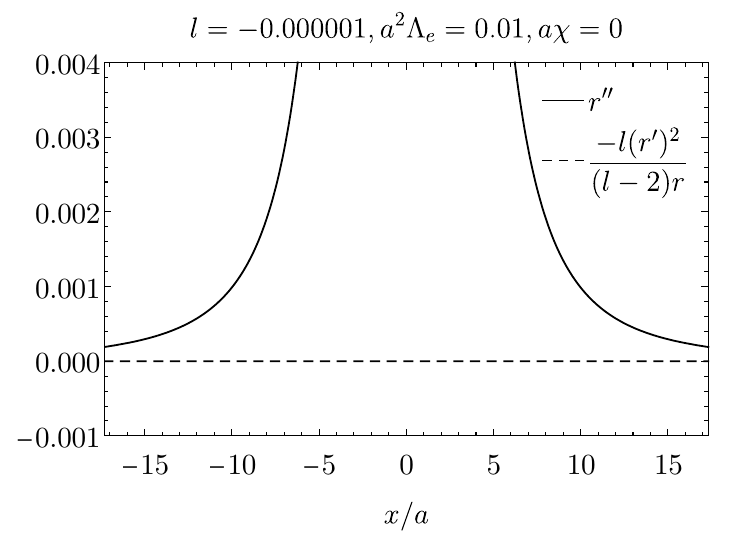}
\includegraphics[width=\columnwidth]{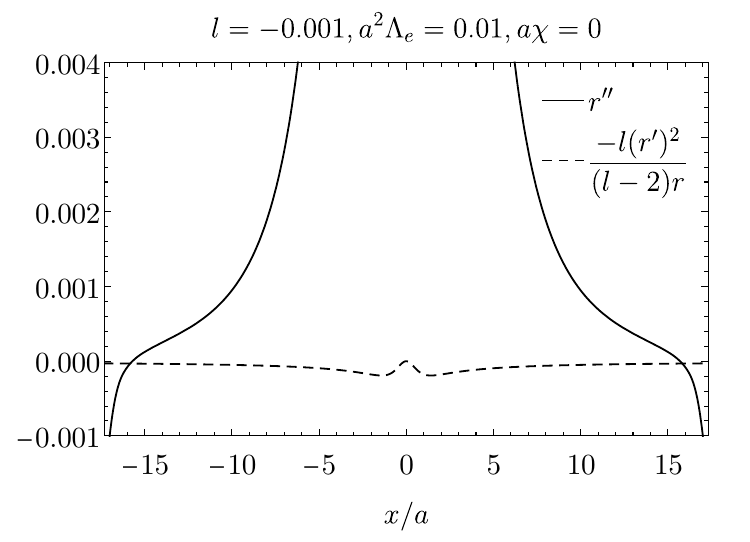}
\includegraphics[width=\columnwidth]{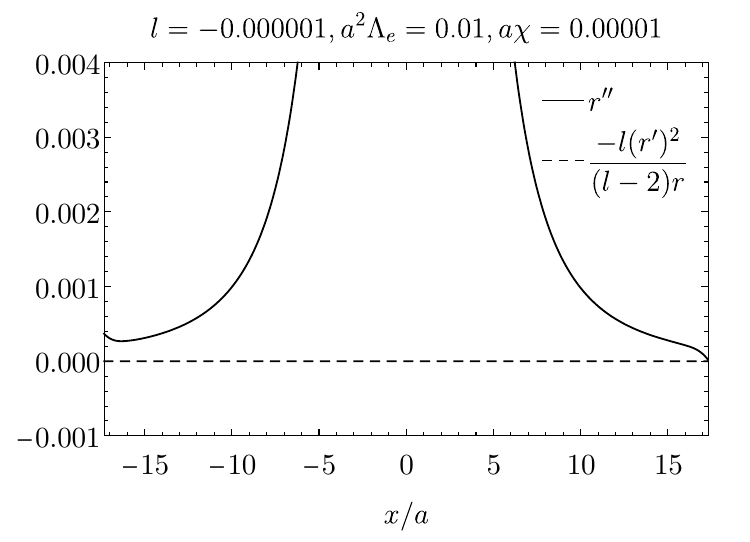}
\includegraphics[width=\columnwidth]{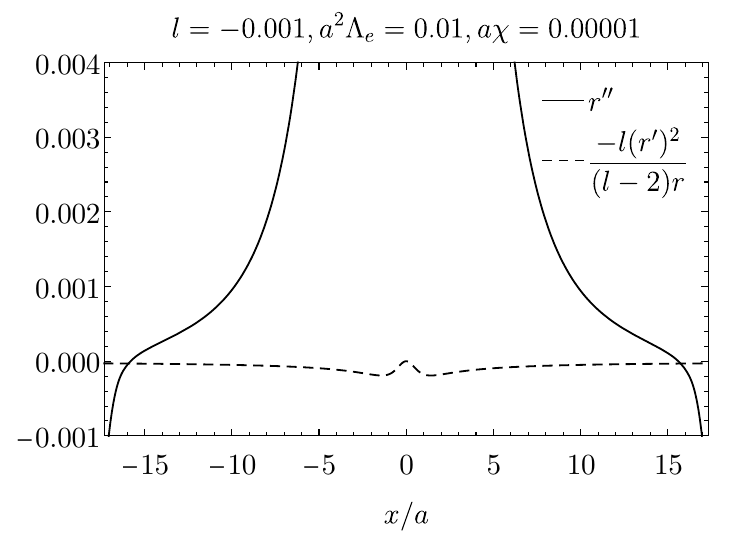}
\includegraphics[width=\columnwidth]{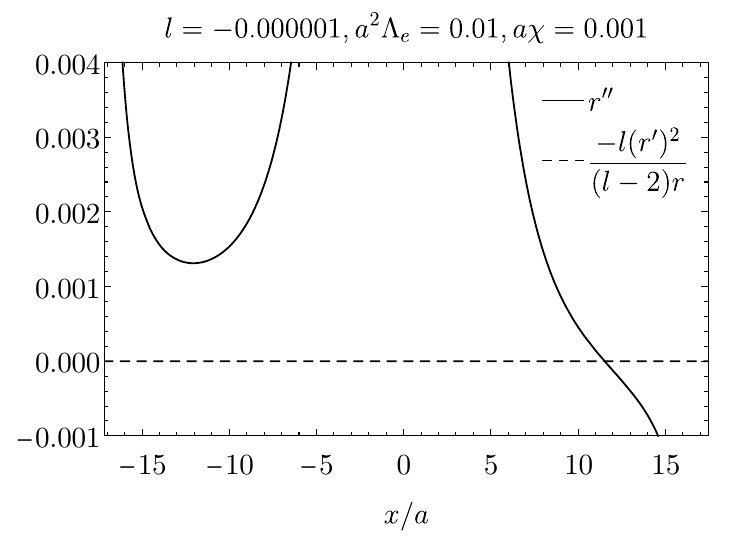}
\includegraphics[width=\columnwidth]{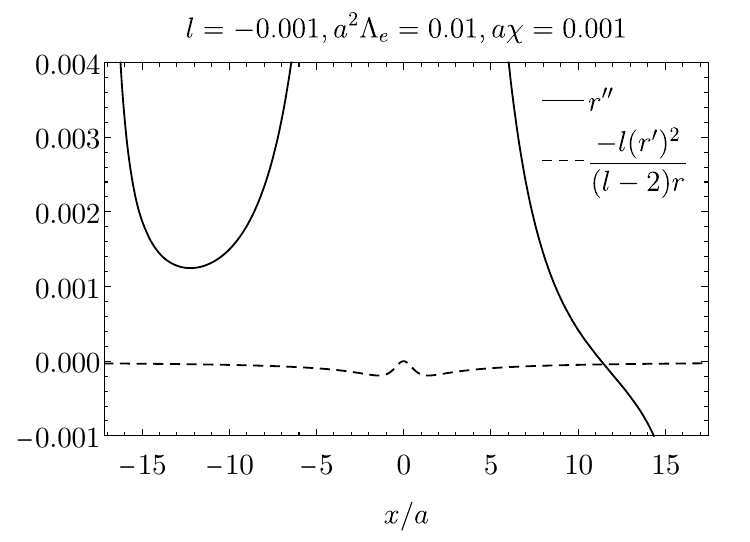}
\caption{Condition for the realness of the scalar field of LV wormholes with positive cosmological constant.}
\label{fig:ds_scalar_field_cond}
\end{figure*}

Remarkably, the presence of the positive effective cosmological constant enables wormholes to be supported by real scalar fields. To see this, one verifies the condition~\eqref{eq:cond_field}. We show in Fig.~\ref{fig:ds_scalar_field_cond} different scenarios.

\section{The anatomy of Lorentz-violating wormholes}~\label{secv}
In the previous sections we have derived four classes of LV wormholes with distinct asymptotic regions, that can be generally represented by the line element~\eqref{metric}, explicitly
\begin{equation*}
    ds^2=-A(x)dt^2+\frac{1}{A(x)}dx^2+r^2(x)d\Omega^2.
\end{equation*}
In summary, as a consequence of a quadratic potential driving the breaking of the Lorentz symmetry, the redshift function, $A(x)$, is constrained to be a linear function of the radial coordinate $x$. It leads to two classes of LV wormholes, namely
\begin{itemize}
    \item[(\textit{i})] \textit{Lorentz-violating Ellis-Bronnikov wormholes};
    \item[(\textit{ii})] \textit{accelerated Lorentz-violating wormholes}.
\end{itemize}
While, by considering a linear potential driving the breaking of the Lorentz symmetry it leads to two another classes of LV wormholes, namely
\begin{itemize}
    \item[(\textit{iii})] \textit{AdS-like Lorentz-violating wormholes};
    \item[(\textit{iv})] \textit{dS-like Lorentz-violating wormholes}.
\end{itemize}
The aim of this section is to gain some understanding on the structure of these wormholes. 

\subsection{Embedding diagrams}
\begin{figure*}
    \centering
    \includegraphics[width=\columnwidth]{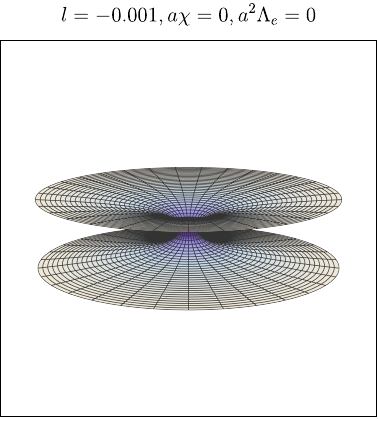}\includegraphics[width=\columnwidth]{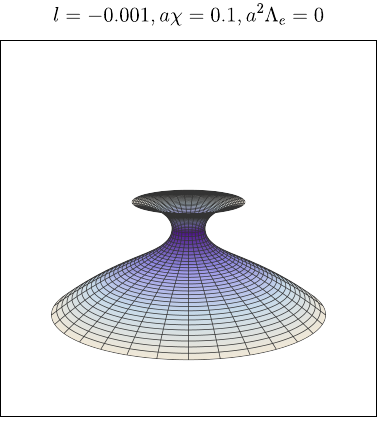}
    \includegraphics[width=\columnwidth]{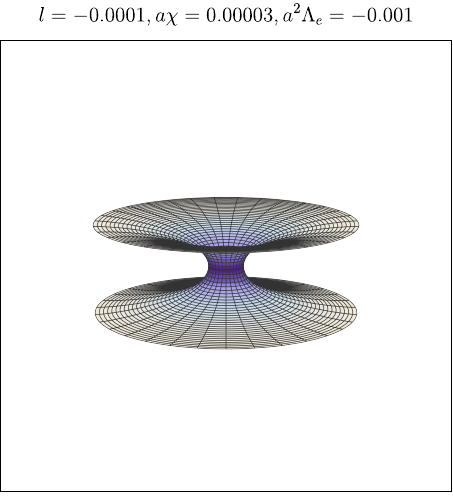}\includegraphics[width=\columnwidth]{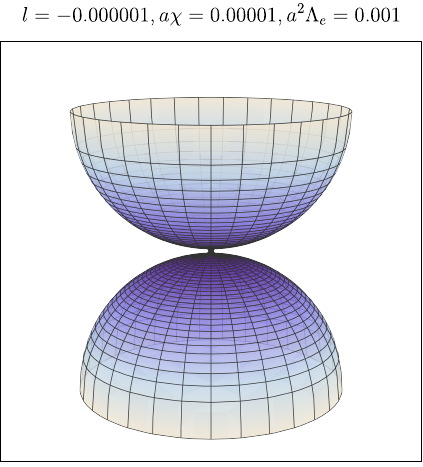}
    \caption{Embedding diagram of the four discussed types of LV wormholes. Top left panel: embedding diagram of a LVEB wormhole. Top right panel: embedding diagram of a ALV wormhole. Bottom left panel: embedding diagram of a AdSLV wormhole. Bottom right panel: embedding diagram of a dSLV wormhole.}
    \label{fig:embedding}
\end{figure*}

Both the redshift function, $A(x)$, and the areal function, $r(x)$, significantly changes the structure of the LV wormholes. One can properly visualize it by drawing the embedding diagrams of the four LV wormholes obtained. We consider, at a given time coordinate $t$, the equatorial plane $\theta=\pi/2$, where the metric~\eqref{metric} reduces to $ds^2=(1/A(x))dx^2+r^2(x)d\phi^2$. One may embed this hypersurface in the three-dimensional Euclidean space $ds^2 = dz^2 + d\rho^2 + \rho^2d\phi^2$. The polar radius can be identified as the radial function, i.e. $\rho=r(x)$, yielding
\begin{equation}
    \label{eq:embedding}
    (z')^2 = \frac{1}{A}-(r')^2,
\end{equation}
which can be solved for $z(x)$. In order to properly construct the embedding of the 2D surface in $\mathbb{E}^3$, the right hand side of Eq.~\eqref{eq:embedding} must be non-negative, therefore $1/A(x)\geq (r')^2$. We show in Fig.~\ref{fig:embedding} an embedding diagram of each type of LV wormhole discussed in the last sections. Remarkably, the embedding diagram of equatorial plane of the dS-like LV wormhole, depicted in the right bottom panel, is fully contained in the Euclidean space up to its cosmological horizons. However, as we can also see in the Fig.~\ref{fig:embedding}, it is not, in general, guaranteed that the the whole 2D hypersurface can be embedded in $\mathbb{E}^3$. The AdS-like LV wormhole, for example, can only be embedded near the throat, while the ALV can be embedded from a region near the throat to the Rindler horizon. We point out that, the LVEB can always be embedded in the Euclidian space if $l\leq 0$. To see this, we compute the condition $1/A(x)\geq (r')^2$ to the LVEB wormhole, obtaining
\begin{equation}
    \frac{l}{-1+l}+\frac{1}{1-l+x^2/a^2}\geq 0.
\end{equation}
This is true for any value of $x$ only if $l\leq 0$. Additionally, far from the throat, $(z')^2$ goes to a constant value, namely $l/(l-1)$, highlighting the non-flat character of the asymptotic region.  

\begin{figure}
    \centering
    \includegraphics[width=\columnwidth]{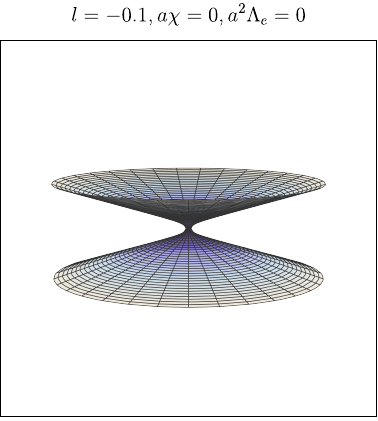}
    \caption{Embedding diagram of 2D section of a LVEB wormhole.}
    \label{fig:embedding_conical}
\end{figure}

It is worth noting that, the asymptotic region of the LVEB wormhole has a similar asymptotic of spacetimes with topological defects, specifically the ones with global monopoles~\cite{barriolaGravitationalFieldGlobal1989} or the ones with cloud of strings~\cite{barbosaRotatingLetelierSpacetime2016}. In Fig.~\ref{fig:embedding_conical} we show the isometric embedding of a 2D slice of a LVEB wormhole with $l=-0.1$, where one notice the cone geometry of the equatorial plane far from the throat, similarly to the behavior of the embedding diagrams of wormholes with a global monopole charge~\cite{jusufi2018conical,sarkar2020traversable}.

Asymptotically, the line element of the LVEB wormhole becomes
\begin{equation}
    \label{eq:asymptotic_LVEB} ds^2 = -dt^2+dx^2+\frac{x^2}{1-l}d\Omega^2.
\end{equation}
Such line element can be put in a more familiar form
\begin{equation}
    \label{eq:asymptotic_LVEB} ds^2 = -dt^2+dx^2+(1-\kappa \eta^2)x^2d\Omega^2,
\end{equation}
where $\eta$ 
is related to the LV parameter via $\kappa \eta^2 = l/(l-1)$. Hence, the cone geometry of the equatorial plane could be anticipated from this identification. Additionally, the cone geometry has a deficit/excess angle $\delta=\pi \kappa \eta^2=\pi l/(l-1)$. Since this asymptotic behavior is present in other LV solutions~\footnote{While the vacuum solutions that consider a background antisymmetric rank-2 tensor yield the same asymptotic behavior shown here~\cite{Yang:2023wtu,Liu:2024oas}, vacuum solutions of Einstein-bumblebee gravity have a slightly different asymptotic~\cite{Casana:2017jkc}, namely $ds^2=-dt^2+dr^2+[1/(1+\ell)]r^2 d\Omega^2$, where $\ell$ is associated to the VEV of the bumblebee field.}, 
it is expected that this angle deficit/excess -- originating from some topological defect -- is a consequence of the LV. Further investigations on how the spontaneous breaking of the Lorentz symmetry give rise to such topological defect will be addressed in a forthcoming work. 

\subsection{Curvature invariants}
In order to gain further understanding on the solutions we have found, we investigate some curvature invariants of these solutions. The Ricci and the Kretschmann scalars of the metric (\ref{metric}) are, respectively, given by
\begin{align*}
    R =&-\frac{r^2 A''+4 r \left(A' r'+A r''\right)+2 A \left(r'\right)^2-2}{r^2} \numberthis\\
    K = &\frac{r^4 \left(A''\right)^2+2 r^2 \left(A' r'+2 A r''\right)^2+2 r^2 \left(A'\right)^2 \left(r'\right)^2}{r^4}\\
&+\frac{4 \left(A \left(r'\right)^2-1\right)^2}{r^4}.\numberthis
\end{align*}
At the throat, the curvature scalars of the LV wormholes are finite. Specifically, one can compute for the LVEB wormhole, namely
\begin{align*}
    \lim_{x\to0}R_{LVEB} =&\frac{2 (l+1)}{a^2 (l-1)} \numberthis\\
    \lim_{x\to0}K_{LVEB} = &\frac{4}{a^4} \left(\frac{2}{(l-1)^2} + 1 \right),\numberthis
\end{align*}
for the ALV wormhole, the same behavior as the LVEB is found at the throat, namely
\begin{align*}
    \lim_{x\to0}R_{ALV} =&\frac{2 (l+1)}{a^2 (l-1)} \numberthis\\
    \lim_{x\to0}K_{ALV} = &\frac{4}{a^4} \left(\frac{2}{(l-1)^2} + 1 \right),\numberthis
\end{align*}
remarkably this limit is independent of $\chi$. For the (A)dSLV wormhole one finds
\begin{align*}
    \lim_{x\to0}R_{(A)dSLV} =&\frac{6(l+1)+2a^2(1-6l)\Lambda_e}{3a^2 (l-1)} \numberthis\\
    \lim_{x\to0}K_{(A)dSLV} = &\frac{4}{a^4} \left(1+\frac{a^4\Lambda_e^2}{9}+\frac{(6+a^2(2-7l)\Lambda_e)^2}{18(l-1)^2}\right).\numberthis
\end{align*}

In the limit LV vanishes, (\(l\to0\)), the curvature invariants of the LVEB and ALV wormholes, at the throat, reduce to ones of the EB wormholes, i.e.,  $R_{EB}|_{x=0} = -2/a^2$ and $K_{EB}|_{x=0} = 12/a^4$. On the other hand, the invariants of the (A)dSLV wormholes still have a non-trivial dependence on $\Lambda_e$.

Far from the throat, the leading terms of the scalars of the LVEB wormhole are 
\begin{align*}
    R_{LVEB} \approx \,&\frac{2 l^2}{x^4}+ \frac{4 a^2 (l-1) l^2}{x^6}+\frac{2 a^4 l (l (3 (l-2) l+5)-4)}{x^8} \\
    &+\frac{4 a^4}{x^8}+\mathcal{O}\left(\frac{1}{x^{10}}\right)\numberthis\\
     K_{LVEB} \approx \, &\frac{4 l^2}{x^4} + \frac{8 a^2 l \left(l^2-1\right)}{x^6} + \frac{12 a^4 l^2 \left(l^2-2\right)}{x^8}\\&+\frac{12 a^4}{x^8}+\mathcal{O}\left(\frac{1}{x^{10}}\right)\numberthis
\end{align*}
As we can see the LV contribution has a slower fall than the terms independent of $l$. This is a consequence of the non-flat asymptotic it presents.

Regarding the ALV wormhole solution~\eqref{eq:ALV_wormhole}, the spacetime is asymmetric and presents a Killing horizon in one of the wormhole sides. Therefore, it is crucial to analyze the curvature invariants far from the throat as well as at the horizon. At the horizon, if $1<N_1<2$, the Ricci scalar is finite and given by
\begin{equation}
    \lim_{x\to-\frac{1}{\chi}}R_{ALV} = \frac{2 N_1 (5 N_1-3) \chi ^2}{(N_1-1) \left(N_1 \left(a^2 \chi ^2+4\right)-2\right)}.
\end{equation}
Conversely, if $N_1<1$ the Ricci scalar diverges as $x\to-1/\chi$. Hence, the Killing horizon is singular, and there is a naked curvature singularity in one side of the wormhole. By computing the Kretschmann scalar at the horizon, we find that it is finite if $3/2<N_1<2$ and given by
\begin{equation}
    \lim_{x\to-\frac{1}{\chi}}K_{ALV} =\frac{4 N_1^2 \left(5 N_1^2-6 N_1+2\right) \chi ^4}{(N_1-1)^2 \left(N_1 \left(a^2 \chi ^2+4\right)-2\right)^2}.
\end{equation}
As a consequence, ALV wormholes are non-singular only for $l>0$ values of the LV parameter. Similarly as the LVEB wormhole case, far from the throat the invariants of the ALV wormhole fall off to zero, however since the expression of these curvature invariants is intricate, we do not show it here. 

\begin{figure*}
    \includegraphics[width=\columnwidth]{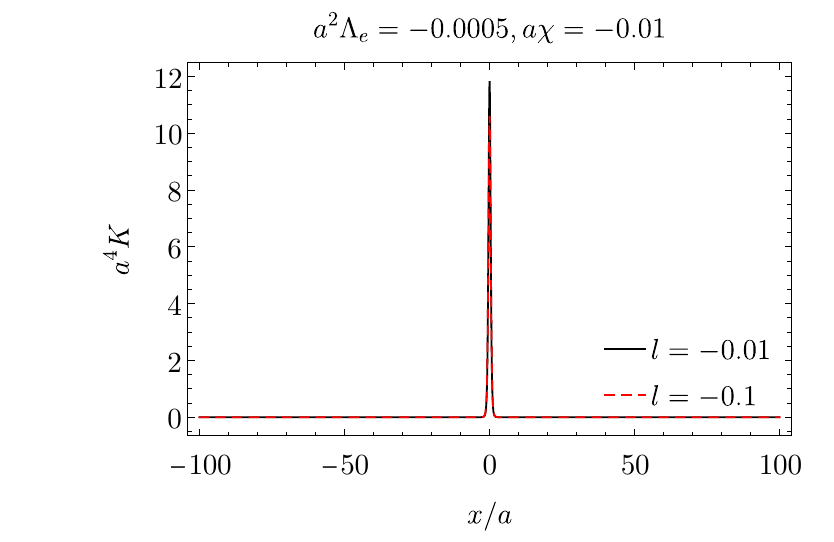}\includegraphics[width=\columnwidth]{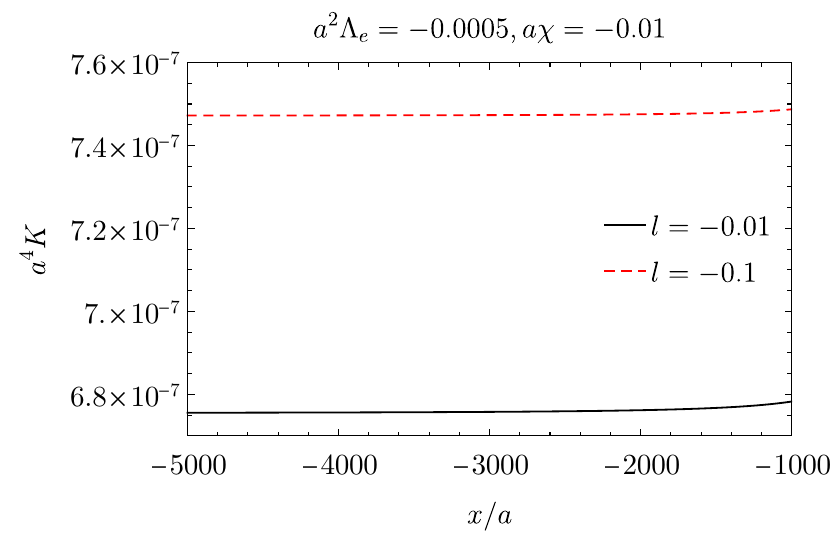}
    \includegraphics[width=\columnwidth]{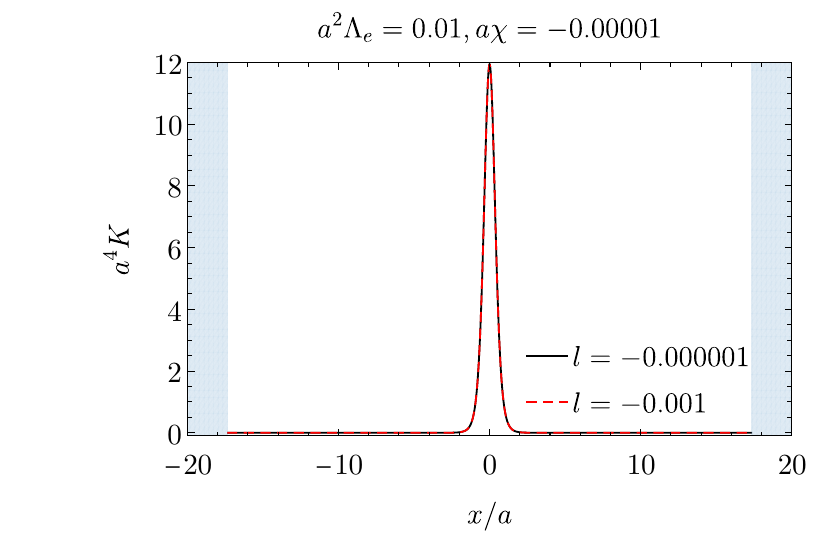}\includegraphics[width=\columnwidth]{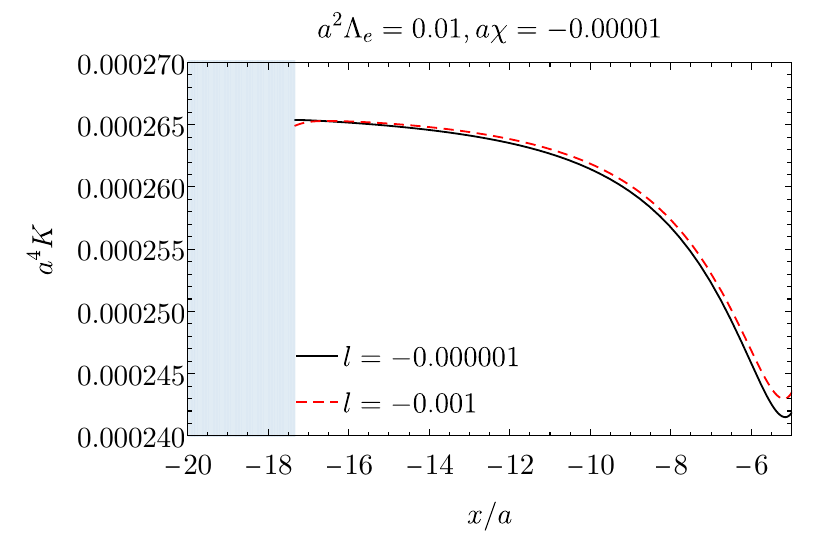}
    \caption{Kretschmann scalar of some (A)dSLV wormholes. In the top row, we depict two AdSLV wormholes for two different values of LV parameter. In bottom row, we depict two dSLV wormholes for two different values of LV parameter. The blue shaded regions are the regions beyond the cosmological horizons.}
    \label{fig:CI_(A)dS}
\end{figure*}

A detailed analysis of the asymptotic behavior of the curvature invariants of (A)dSLV wormholes is more difficult since we do not have access to the analytical expression of the areal radius function. However, by using the numerical results of the last section we can plot the curvature invariants and gain some insights of the asymptotic limits. We show in Fig.~\ref{fig:CI_(A)dS} the Kretschmann scalar for some (A)dSLV wormholes. These examples illustrate that, when a cosmological constant is considered through the inclusion of a linear potential driving the breaking of the Lorentz symmetry, the spacetime of (A)dSLV wormholes may be free of curvature singularities.

\subsection{Light paths}
In this subsection we investigate potential observational signatures associated with the trajectory of light in LV wormhole spacetimes. The Lagrangian of a particle in the background of LV wormholes is
\begin{equation}
    2{\mathcal{L}} = -A(x)\dot{t}^2+\frac{\dot{x}^2}{A(x)}+r^2(x)(\dot{\theta}^2+\sin^2\theta\dot{\varphi}^2)=\varepsilon,
\end{equation}
where $\varepsilon=-1$ for massive particles and $\varepsilon=0$ for massless particles. Since $t$ and $\varphi$ are cyclic coordinates, there are two quantities conserved along the geodesics, namely
\begin{align}
    E&=A(x)\dot{t},\\
    L&=r^2(x)\sin^2\theta \dot{\varphi},
\end{align}
usually associated, respectively, with the energy and angular momentum of the particle but since this is spacetime is not asymptotically flat, such interpretation is not that direct. 
At the equatorial plane ($\theta=\pi/2$), one obtains the orbit equation
\begin{equation}
    \dot{x}^2=U(x)=E^2-A(x)\left(\frac{L^2}{r^2(x)}-\varepsilon\right),
\end{equation}
where $U(x)$ is the effective potential. A critical point of the potential, obtained through $U'(x_c)=0$, or explicitly in terms of $A$ and $r$,
\begin{equation}
    \label{eq:LR}r(x_c)A'(x_c)-2A(x_c)r'(x_c)=0,
\end{equation}
locates an unstable circular orbit if $U''(x_c)>0$ and a stable circular orbit if $U''(x_c)<0$. 

Light rays, $\varepsilon=0$, impinging from infinity with impact parameter $b=L/E$, can reach a closest approach $x_0$ before be scattered back to infinity with a total deflection angle
\begin{equation}
   \label{eq:deflection} \Omega(x_0)=2\int_{x_0}^\infty\frac{A(x)r(x_0)dx}{\sqrt{A(x_0)r^2(x)-A(x)r^2(x_0)}}-\pi,
\end{equation}
where closest approach is calculated from $\sqrt{r^2(x_0)/A(x_0)}=b$. In particular, if a light ring is present (planar circular light-like geodesics), it is associated with a critical impact parameter $b_c=\sqrt{r^2(x_c)/A(x_c)}$. If $b<b_c$ the light ray crosses the throat and propagates along the other side of the wormhole. As $x_0\to x_c$, $\Omega(x_c)$ diverges logarithmically.

\subsubsection{Light rings}
From Eq.~\eqref{eq:LR}, one finds that the EB wormhole has a single light ring in the spacetime, located at $x=x_c=0$. Therefore, the only light ring of an EB wormhole is at the throat. It is worth investigate how is the light ring structure of LV wormholes, and how it tells apart from the EB one. The simplest lapse function of the LVEB wormhole implies that
\begin{equation}
    2r'(x_c)= \frac{x_c}{(1-l)\sqrt{a^2+x_c^2/(1-l)}}=0.
\end{equation}
Therefore, in the LVEB wormhole there is also a single light ring, and it is located at the throat. 

On the other hand, by considering the ALV wormhole, Eq.~\eqref{eq:LR} becomes 
\begin{equation}
    \frac{-2+4N_1+a^2N_1\chi^2-2(2N_1-1)(1+\chi x_c)^{N_1}}{N_1\sqrt{a^2+\frac{2(2N_1-1)(-1-N_1 x_c \chi +(1+x_c\chi)^{N_1})}{(N_1-1)N_1\chi^2}}}=0,
\end{equation}
that can be solved for $x_c$, where one gets
\begin{equation}
    x_c=\frac{-1+\left(1+\frac{a^2N_1\chi^2}{4N_1-2}\right)^{1/N_1}}{\chi}.
\end{equation}
Therefore, ALV wormholes also have a single light ring, but it is shifted from the throat. Such shifting is due to the asymmetry introduced from the Rindler-type acceleration. In the limit $\chi\to 0$, an expansion of the above expression shows that $x_c\to 0$, as expected.

Now, by considering (A)dSLV wormholes, light rings are found as roots of
\begin{equation}
    \left(\chi -\frac{2\Lambda_e x_c}{3}\right)r(x_c)+\frac{2}{3}(-3-3\chi x_c+\Lambda_e x_c^2)r'(x_c)=0.
\end{equation}
As we can see, if $\chi\neq 0$, the throat, $x=0$, cannot support a light ring. If $\chi=0$, the spacetime is symmetric, and expected from the result of Ref.~\cite{xavierTraversableWormholesLight2024}, there is at least one light ring in the spacetime, and it is located the throat. We point out that, our numerical searches do not show any new light ring in the spacetime other than the one located at the throat.

\subsubsection{Deflection of light}
Before we finish this section, let us investigate the deflection of light by LVEB wormholes. In terms of the impact parameter of the photon, Eq.~\eqref{eq:deflection} becomes, for the LVEB wormhole,
\begin{equation}
    \Omega(b)=2\int_{x_0(b)}^\infty \frac{b dx}{\sqrt{r^2(x)(r^2(x)-b^2)}}-\pi.
\end{equation}
By performing the variable transformation $z=b/r(x)$, the above expression becomes
\begin{equation}
    \Omega(b)=2\sqrt{1-l}K(z)-\pi,
\end{equation}
where $K(z)$ is the complete elliptic integral of the first kind, given by
\begin{equation}
    K(z) \equiv \int_{0}^1 \frac{dz}{\sqrt{1-z^2}\sqrt{1-k^2z^2}},
\end{equation}
and $k=a/b$. Under the weak-field approximation, $a\ll b$, the deflection angle can be approximated by
\begin{align*}
    \Omega(b)\approx &\,\pi\left(\sqrt{1-l}-1\right)+\frac{\pi\sqrt{1-l}}{4}\frac{a^2}{b^2}\\&+\frac{9\pi\sqrt{1-l}}{64}\frac{a^4}{b^4}+\mathcal{O}\left(\frac{a^6}{b^6}\right),\numberthis
\end{align*}
that, in the vanishing of the LV parameter, reduces to the well-known result for EB wormholes~\cite{nakajimaDeflectionAngleLight2012,tsukamotoStrongDeflectionLimit2016}. As we can see, the LV parameter \textit{spoils} the total deflection angle, even in the weak-field approximation. This is due to the angle deficit/excess in the spacetime introduced by the LV. 

\section{Conclusion}\label{con}
We have obtained traversable wormholes supported by phantom scalar fields in a Lorentz-violating gravity framework. The field that carries the effects of LV is an antisymmetric rank-2 tensor with a non-zero VEV. Such tensor field is non-minimally coupled to the Riemann curvature tensor, which proved to be a suitable scenario to look for wormhole solutions in the presence of LV. The reason behind it lies in the mild constraints imposed by the LV on the areal radius. The VEV of the antisymmetric rank-2 tensor, nonetheless, imposes constraints on the lapse function. As a consequence, under the VEV configuration adopted, and in the absence of couplings of the background tensor to the matter fields, the allowed lapse functions are either constant, linear or quadratic, depending on the self-interaction potential that drives the breaking of the Lorentz symmetry.

Regarding the quadratic potentials, we have found analytical solutions of wormholes with constant and linear lapse functions. The former, called LVEB wormholes, reduce to the symmetric EB wormhole if the LV is turned off. Such wormholes present a non-flat asymptotic, with their equatorial plane presenting a cone geometry with deficit/excess angle $\delta=\pi l/(l-1)$. The effect of this deficit/excess angle manifests even in the weak-field approximation, as can be seen in the leading-order terms of the deflection angle of light. Since the lapse function in this case is a constant, the only light ring present in the spacetime is an unstable one at the throat. The other possibility, allowed by the quadratic potential, is a lapse function linear on the radial coordinate. This linear term depends on a parameter $\chi$, that can be interpreted as a Rindler-type acceleration. Due to it, such wormholes are called ALV wormholes. The parameter $\chi$ introduces a Killing horizon in one side of the wormhole, hence it implies in an asymmetry in the spacetime. This asymmetry forces the only light ring of the sapcetime to shift to one side of the throat. The drawback of this solution is that, regardless of the LV parameter, the scalar field that supports the wormhole unavoidably suffers a transition from a real to a complex scalar field from some ranges of the radial coordinate. Such transition may indicate instabilities in such spacetimes, and a detailed investigation on this subject should be done.

Concerning the linear potentials, we have found numerical solutions of wormholes with quadratic lapse functions. The magnitude of the quadratic term is controlled by the Lagrange multiplier field through an effective cosmological constant $\Lambda_e=3\kappa\lambda/\xi$. These quadratic dependencies are commonly associated to (A)dS asymptotic, and due to it we identify two classes of wormholes, namely AdSLV and dSLV wormholes, depending on the sign of the parameter $\Lambda_e$. We point out that, linear terms controlled by a Rindler-type acceleration may also appear in the allowed lapse function, introducing, therefore, an asymmetry in the spacetime. Remarkably, in the presence of an effective cosmological constant, the scalar field that supports the wormhole can be a real-valued function in the whole spacetime. Hence, it does not present any transition from a real-valued to a complex-valued function. It is important to remark out that the spherical sector of these wormholes, even though similar the one of (A)dS spacetimes, does not match with them neither far from the throat. Such difference suggests that the LV modifies non-trivially the large structure of the spacetime, and an in-depth investigation on this matter should be conducted.

Our findings are highly influenced by the VEV of the antisymmetric rank-2 tensor we have chosen. Different configurations should allow for less restrictive lapse functions. A systematic exploration on different vacuum configurations should be performed in order to unveil new compact objects supported by LV theories, which include, not only novel wormholes, but also black holes and black bounces. Another perspective is the inclusion of couplings subjected to LV on the matter fields. This should allow for simple wormhole configurations, such as the LVEB wormhole investigated here, in Lorentz-violating gravity with non-minimal couplings to the Ricci tensor, such as the bumblebee model~\cite{Kostelecky:2003fs}. Furthermore, it is of interest to the authors to explore the observational signatures of these solutions in different contexts, including shadow analysis and the quasi-normal mode spectrum.  A careful analysis of these potential scenarios is underway.
\begin{acknowledgments}
The authors would like to acknowledge Fundação de Amparo à Pesquisa e ao Desenvolvimento Científico e Tecnológico do Maranhão (FAPEMA),  Conselho Nacional de Desenvolvimento Cient\'ifico e Tecnol\'ogico (CNPq), Coordena\c{c}\~ao de Aperfei\c{c}oamento de Pessoal de N\'ivel Superior (CAPES) -- Finance Code 001, from Brazil, for partial financial support. R.B.M. is supported by CNPq/PDJ 151250/2024-3.  L.A.L is supported by FAPEMA BPD- 08975/24.
\end{acknowledgments}

\bibliography{refs.bib}
\bibliographystyle{report}
\end{document}